\documentclass[structabstract]{aa}  

\usepackage{txfonts}
\usepackage{graphicx}

\begin{document}

\title{The star formation histories of Hickson compact group galaxies}

\author{I.~Plauchu-Frayn\inst{1} \and A.~Del Olmo\inst{1}
\and R.~Coziol\inst{2} \and J.~P.~Torres-Papaqui\inst{2}
} 

\institute{Instituto de Astrof\'isica de Andaluc\'ia IAA -- CSIC, 
Glorieta de la Astronom\'ia, s/n, E-18008 Granada, Spain\\  
\email{ilse@iaa.es}
\and 
Departamento de Astronom\'ia, Universidad de Guanajuato, 
Apdo. Postal 144, 36000, Guanajuato, Gto., Mexico
}

\date{Received date; Accepted date}

\abstract
{}
{
We study the Star Formation History (SFH) of 210 galaxies members of 55 Hickson Compact Groups 
(HCG) and 309 galaxies from the Catalog of Isolated Galaxies (CIG).  The SFH traces the variation of star formation 
over the lifetime of a galaxy, and yields consequently a snapshot picture of its formation. Comparing the SFHs in 
these extremes in galaxy density, allows us to determine the main effects of Compact Groups (CG) on the formation of 
galaxies.
}
{
We fit our spectra using the spectral synthesis code STARLIGHT and obtain the stellar population contents and mean 
stellar ages of HCG and CIG galaxies in three different morphological classes: early-type galaxies (EtG), 
early-type spirals (EtS), and late-type spirals (LtS). 
}
{
We find that EtG and EtS galaxies in HCG show larger contents of old and intermediate stellar populations 
as well as an important deficit of the young stellar population, what clearly implies an older average stellar age 
in early galaxies in HCG. For LtS galaxies we find similar mean values for the stellar content and age in the
two samples. However, we note that LtS can be split into two subclasses, namely old and young LtS. In HCG 
we find a higher fraction of young LtS than in the CIG sample, in addition, most of these galaxies belong to 
groups where most of the galaxies are also young and actively forming stars.
The Specific Star Formation Rate (SSFR) of spiral galaxies in the two samples differ. EtS in HCG show lower values 
of the SSFR, while LtS peak at higher values when comparing with their counterparts in isolation.
We have also measured shorter Star Formation Time Scale (SFTS) in HCG galaxies, indicating that they have less 
prolonged star formation activity than CIG galaxies. We take these observations as evidence that galaxies in 
CG evolved more rapidly than galaxies in isolation regardless of their morphology.
Our observations are consistent with the hierarchical galaxy formation model, which states that CG are 
structures that formed recently from primordial small mass density fluctuations. From the systematic difference 
in SFTS we deduce that the HCG have most probably formed $\sim 3$~Gyr in the past. The galaxies in the HCG are not in  
equilibrium state, but merging without gas (i.e., under dissipationless conditions), which may explain their relatively 
long lifetime.
}
{}


\titlerunning{Star formation histories of HCG galaxies}

\maketitle

\section{Introduction}

Compact Groups (CG) of galaxies are local structures with high galaxy densities, equivalent to what is measured in the
centers of clusters of galaxies. Contrary to clusters, however, the velocity dispersion of the galaxies in CG is low,
comparable to the velocity dispersion of elliptical galaxies or the rotational velocity of spiral disks (Hickson \cite{Hickson97}
and references therein). These conditions favor interactions, and even mergers, which suggests that galaxies in CG 
must show plenty evidence of these phenomena, happening now, or that happened in the past. However, after almost forty 
years of studies, determining how CG really formed still remains an open question.

One of the most extensively studied samples of CG is the Hickson Compact Group (HCG) catalog as defined by Hickson 
(\cite{Hickson82}). Subsequent spectroscopy studies have shown that the majority of the HCG have galaxies with 
concordant redshifts (Hickson et al. \cite{Hickson92}), that display many signs of interactions, under the form of 
tidal tails, tidal bridges, distorted isophotes, faint shells, and double nuclei (Mendes de Oliveira \& Hickson 
\cite{Mendes94}; Hickson \cite{Hickson97}; V{\'{\i}}lchez  \& Iglesias-P{\'a}ramo \cite{Vilchez98}; Verdes-Montenegro 
et al. \cite{Verdes01}, \cite{Verdes02}, \cite{Verdes05}; Coziol \& Plauchu-Frayn \cite{Coziol07}; Plauchu-Frayn \& 
Coziol \cite{Plauchu10a}). Asymmetrical rotation curves were also detected (Rubin et al. \cite{Rubin91}; Nishiura et al. 
\cite{Nishiura00}; Torres-Flores et al. \cite{Torres10}) and, contrary to what was assumed in the past, many evidence 
for present and past mergers were also clearly established (Caon et al. \cite{Caon94}; Coziol \& Plauchu-Frayn 
\cite{Coziol07}; Plauchu-Frayn \& Coziol \cite{Plauchu10b}). All these observations are consistent with the initial 
assumption which sustains that CG are physically real structures, where  galaxies during their formation and evolution 
have experienced frequent dynamical interactions. 

Other aspects of CG, on the other hand, may still defy our comprehension. Spectroscopy study results, in particular, were 
found to be different than expectations based on the common assumptions that the main effects of dynamical interactions 
on galaxies in CG should be a direct enhancement of their star formation and activation of supermassive black holes (SMBHs) 
at their centers. Although emission-line galaxies are remarkably frequent in CG, involving more than 50\% of the galaxies 
(Coziol et al. \cite{Coziol98}, \cite{Coziol00}, \cite{Coziol04}; Mart\'inez et al. \cite{Martinez08}, \cite{Martinez10}), 
this activity seems mostly nonthermal, under the form of Seyfert~2, LINERs, and numerous low-luminosity active galactic nuclei. 
Luminous AGN with typical broad line regions are relatively rare in CG (Coziol et al. \cite{Coziol98}, \cite{Coziol04}; 
Ribeiro et al. \cite{Ribeiro96}; Mart\'inez et al. \cite{Martinez08}, \cite{Martinez10}). Nuclear star formation, although 
mildly enhanced in some groups (e.g., Ribeiro et al. \cite{Ribeiro96}; Iglesias-P\'aramo \& V\'ilchez \cite{Iglesias97}), 
is generally low (Rubin et al. \cite{Rubin91}; Zepf et al. \cite{Zepf91}; Moles et al. \cite{Moles94}; Pildis et al. 
\cite{Pildis95}; Menon \cite{Menon95}; Ribeiro et al. \cite{Ribeiro96}; Coziol et al. \cite{Coziol98}, \cite{Coziol00}; 
Verdes-Montenegro et al. \cite{Verdes98}; Allam et al. \cite{Allam99}; Iglesias-P{\'a}ramo \& V{\'{\i}}lchez \cite{Iglesias99}; 
Mart\'inez et al. \cite{Martinez10}).

In part, the observation that star formation is possibly deficient and that AGN are mostly underluminous in CG could be
due to a general deficiency of gas. One possibility is that this is the effect of tidal gas stripping. However, although
evidence of tidal gas stripping is unquestionable in CG (e.g., Verdes-Montenegro et al. \cite{Verdes01}; Durbala et al. 
\cite{Durbala08}), this mechanism alone cannot explain all the observations. In particular, tidal gas stripping cannot 
explain the higher fraction of early-type galaxies found in CG (Hickson \cite{Hickson82}; Williams \& Rood \cite{Williams87}; 
Sulentic \cite{Sulentic87}; Hickson, Kindl \& Huchra \cite{Hickson88}; Palumbo et al. \cite{Palumbo95}). Indeed, it is 
difficult to explain how an elliptical galaxy, or even the massive bulge of an early-type spiral galaxy, can build up 
after star formation is quenched in the disk of a late-type spiral galaxy by gas stripping (Dressler \cite{Dressler80}). 
Unless one assumes that these galaxies can build massive bulges later by dry mergers (Plauchu-Frayn \& Coziol 
\cite{Plauchu10b}). However, neither can gas stripping explains the many AGN observed in CG (e.g., Coziol et al. 
\cite{Coziol98}, \cite{Coziol00}; Mart\'inez et al. \cite{Martinez10}). These AGN are related to the formation of numerous 
early-type galaxies through tidal triggering, a mechanism that funnels down gas right into the center of galaxies, starting 
a sequence of bursts of star formation that increase the mass of the bulges, and forming, or feeding, a SMBH at their 
centers (Merritt \cite{Merritt83}; \cite{Merritt84}; Byrd \& Valtonen \cite{Byrd90}; Henriksen \& Byrd \cite{Henriksen96}; 
Fujita \cite{Fujita98}). However, because the level of such activities is generally low, we would have then to assume 
these events have taken place sometime in the recent past, or possibly during the formation of the CG (Coziol et al. 
\cite{Coziol04}; Mendes de Oliveira et al. \cite{Mendes05}). The important question is how long ago?

Recent studies using Lick indices to determine the stellar population contents of elliptical galaxies in small samples of HCG have 
shown that the formation process of these galaxies is similar to that of elliptical in clusters of galaxies (Proctor et al. 
\cite{Proctor04}; Mendes de Oliveira et al. \cite{Mendes05}; de la Rosa et al. \cite{Rosa07}), and suggested their 
formation took place at least $2$~Gyr in the past. This is already older than predicted by the fast merger scenario for 
CG, which have led some authors (e.g, Proctor et al. \cite{Proctor04}; Mendes de Oliveira et al. \cite{Mendes05}) to 
propose that CG are defying the standard cosmological model of structure formation. However, this conclusion is 
somewhat premature, considering our current state of comprehension of the physics behind all the processes entering this model.

Many aspects of CG as described above would rather seem to be in reasonably good agreement with the basic 
predictions of the standard model of structure formation. In Coziol et al. (\cite{Coziol09}) 
it was suggested that CG are the building blocks of larger scale structures like clusters of galaxies. Consistent with 
the hierarchical galaxy formation model (White \& Frenk \cite{White91}; Kauffmann et al. \cite{Kauffmann93}; 
Cole et al. \cite{Cole94}), galaxies form from primordial fluctuations in mass density. In regions showing the higher
fluctuations, the galaxy formation process starts before than regions with lower fluctuations. Assuming the power law 
for the mass function of the fluctuations is relatively steep, the first structures 
that form would look like massive CG, where the low velocity dispersions of the protogalaxies in these systems favor 
interactions and mergers that produce numerous massive elliptical and early-type spiral galaxies through a sequence of 
starbursts. The increased potential wells of these newly assembled structures would then attract other similar 
systems, and a lot of primordial gas would eventually fall in, being enriched in the process by the metals expelled from 
galaxies by starburst winds (Torres-Papaqui et al. \cite{TP12}). Much later, late-type spirals that formed in isolation in 
the field would eventually fall in, making the final structure consistent with a cluster of galaxies. According to 
the hierarchical galaxy formation model, in low mass density fluctuations the same process would be expected to start 
later, resulting, due to the small amounts of mass available, to slightly different results.

Within the above theoretical context our present knowledge of CG may suggest two different scenarios for their formation. According to the first 
scenario galaxies in CG have formed first in relative isolation, from very low mass density fluctuations consistent 
with the field, more likely forming a gas rich disk, then later on (i.e. at the present epoch), started to form a more 
gravitationally bound structure like the CG. Within this view, tidal gas stripping, quenching star formation in disk 
galaxies, and dry mergers, growing more massive bulges, would play the central role in the evolution of CG. The second 
scenario is more similar to what may have happened during the formation of clusters. Galaxies forming the CG never 
passed by a spiral form before, but merged rapidly as gas rich protogalaxy systems. Through multiple interactions, 
triggering starbursts and gas funneling, galaxies form massive bulges, many with an active black hole at their center. 
However, considering the low density of matter where these galaxies formed, and a first phase of intense star formation 
and AGN activity, short time-scales for gas consumption would ensue. Reaching a pure elliptical structure in these 
conditions would not be always possible, and the expected end products would rather be a bunch of gravitationally bounded, 
relatively gas deficient, early-type spiral galaxies, with an aging SMBH at their center accreting at lower rates than 
in the field (consistent with a LLAGN). As in the model for the formation of clusters, during the first phase of the 
CG formation a deeper gravitational potential well builds up, which then act as a pole of attraction for late-type 
galaxies that formed at the periphery, making the final structures similar to CG.

In order to distinguish between the above scenarios for the formation of CG we have used the stellar population
synthesis code \begin{scriptsize}STARLIGHT\end{scriptsize}\footnote{http://www.starlight.ufsc.br} (Cid Fernandes et al.
\cite{Cid05}) to determine the star formation histories (SFHs) of 210 galaxies members in 55 Hickson Compact Groups. 
The SFH traces the variation of star formation over the lifetime of a galaxy, and yields consequently a snapshot picture 
of its formation. For comparison we have also determined the SFHs of 309 galaxies taken from the Catalog of Isolated 
Galaxies (CIG), which represent the lowest galaxy density regime. Our study is organized in the following manner. In 
Sect.~\ref{samples} we describe the selection of our two different samples. In Sect.~\ref{method} we explain how the 
spectra synthesis method works. In Sect.~\ref{results} we present the results of our comparison of the SFHs in our 
different samples. In Sect.~\ref{discussion} we discuss briefly the consequence of these results and expose our main 
conclusions. Throughout this study a Hubble constant $H_0$ = 75 km s$^{-1}$ Mpc$^{-1}$ is assumed.

\section{Selection of the samples and observational data} \label{samples}

\subsection{Compact group galaxies}

To determine our sample of galaxies in CG, we started with the sample of 64 Hickson Compact Groups (HCG; Hickson
\cite{Hickson82}) which was used by Mart\'inez et al. (\cite{Martinez10}, hereafter M10) to determine the nature of the
nuclear activity of the galaxies in these systems. These data comprise medium-resolution long-slit spectroscopy 
in the optical range 3600-7200\AA. They were obtained using four telescopes: the 2.2m telescope in Calar-Alto (CAHA, Spain), 
the 2.56m NOT Telescope at the Observatorio del Roque de los Muchachos (ORM) in La Palma (Spain), the 2.12m telescope 
at the Observatorio Astron\'omico Nacional in San Pedro M\'artir (SPM) in Baja California (M\'exico), and the 1.5m 
telescope operated by the IAA in the Sierra Nevada Observatory (OSN, Spain). The observations were performed using 
a reciprocal dispersion of 2\AA/px at CAHA and OSN, 3\AA/px at the NOT ALFOSC spectrograph and 4.3\AA/px in SPM. 
A detailed description of the instrumental setups, the log of observations, and reduction procedures were presented 
in M10 (Tables 2 and 3) and in Mart\'inez et al. (\cite{Martinez08}). Also included in those references there are the 
characteristics of the final one-dimensional nuclear spectra such as the extracted apertures for each galaxy, that 
in general corresponds to the central 1kpc region.

The total number of galaxies in the 64 HCG amounts to 269, while in M10 we found the spectra of only 220 of these 
galaxies (82\% of the sample). As an effort to complete our sample we searched the Sloan Digital Sky Survey Data 
Release 7 (SDSS-DR7; Abazajian et al. \cite{Abazajian09}) and found the spectra for 68 galaxies. However, 63 of 
these galaxies were already in the M10 sample. Comparing the signal-to-noise ratio, S/N, around three 
different $\lambda$ windows (centered at $\lambda=4750$\AA, $\lambda=5525$\AA, and $\lambda=6825$\AA), we kept 57 spectra from SDSS
and 6 of M10, applying the criterion that S/N$ \ge 5$. We also discarded 15 galaxies from M10 because they were observed
only in the red or because they did not passed our S/N criterion. Our final complete sample consists, therefore, in 210 
galaxies in 55 HCG (148 galaxies from M10 and 62 from SDSS). All the galaxies have concordant redshift (Hickson et al.
\cite{Hickson92}), up to $z=0.045$, and are members of groups having a surface brightness $\mu_g \ge$ 24 mag arcsec$^2$.
All the morphologies for the galaxies in our sample have been extracted from the on-line compilation
HYPERCAT\footnote{http://www-obs.univ-lyon1.fr/hypercat} (Paturel et al. \cite{Paturel94,Paturel97}).

\subsection{Isolated galaxies}

To build our sample of isolated galaxies we have used the Catalog of Isolated Galaxies (CIG; Karachentseva \cite{Karachentseva73}). 
In this catalog, a galaxy with a standard angular diameter $D$ is classified as isolated when all its neighbors, with 
angular diameters $d$, are located at a projected distance not closer than $20d$. Physically, this criterion should 
imply that a galaxy has not suffered any gravitational influence from nearby galaxies over the last billion years. In 
absence of external influence, the evolution of these galaxies is assumed consequently to be driven mainly by internal 
processes.

In their study of isolated galaxies, Verley  et al. (\cite{Verley07}) refined the isolation criterion by taking into 
account not only the local number density of neighbor galaxies, a parameter they identify as $\eta_k$, but also the tidal 
strength per unit mass, $Q$, that is produced by all the neighbors. By applying these criteria these authors have reduced the 
number of CIG by 16\%, producing a list of 791  ``most isolated'' galaxies (defined to have $\eta_k \le 2.4$, and $Q\le -2$). 
Starting with this list and applying our S/N criterion we have extracted from SDSS the spectra of 309 of these galaxies, 
forming our final sample of CIG. The morphologies for these galaxies were extracted from Sulentic et al. (\cite{Sulentic06}) 
and HYPERCAT.

\begin{figure}
\centering
\includegraphics[width=8cm]{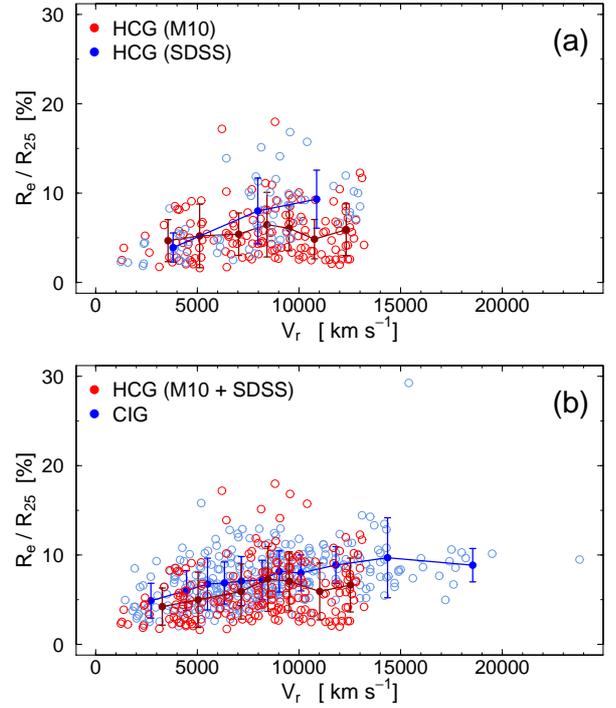}
\caption{Equivalent radius of the used spectrum to radius at 25 mag  arcsec$^{-2}$ isophote ratio as 
a function of radial velocity of galaxies. Panels are: a) comparison between spectra taken from Mart\'inez et al. (\cite{Martinez10}) 
and SDSS for the HCG sample and b) comparison between the HCG and CIG samples. In each panel, the lines join the mean values 
determined in sets of: a) 20  and b) 30 galaxies at different redshifts. The error bars give the standard 
deviation of the means.} 
\label{aperture}
\end{figure}

\subsection{Quantifying aperture biases}

Because the spectra come from two different database and have been taken with different instruments, we have 
verified that they are comparable and suitable for our study. The spectral resolution of the spectra, which are both 
in the medium range, is not a problem. Both have the value required to be adjustable to the stellar library resolution 
used by \begin{scriptsize}STARLIGHT\end{scriptsize}. On the other hand, the SDSS spectra have been taken with a 
fixed fiber circular aperture while the M10 spectra have been taken with a long slit aperture.

For compatibility sake, we have checked to see if 
the extracted spectra cover similar physical spatial areas. To do so, we have calculated the equivalent radius, $R_e$, 
covered by the long-slit and fiber apertures and compared these values with the radius $R_{25}$, which corresponds in a 
galaxy to the radius of the 25 mag arcsec$^{-2}$ isophote. In Fig.~\ref{aperture} we compare the percentage of $R_{25}$ 
covered by $R_e$ in our samples. In our HCG sample, the mean value of $R_e/R_{25}$ is $(5\pm 3)\%$ and $(7\pm 4)\%$ 
for spectra taken from M10 and SDSS, respectively, while in the CIG sample this is $(7\pm 3)\%$. In Fig.~\ref{aperture}a we plot 
$R_e/R_{25}$ for galaxies in M10 and SDSS for HCG sample and in Fig.~\ref{aperture}b the same comparison is shown between 
the HCG and CIG samples. In this figure, the continuous lines join the mean values determined in sets of 20 and 30 galaxies 
taken at different redshifts (Fig.~\ref{aperture}a and b, respectively), and the error bars give the standard deviations 
of the means. The means are comparable to the medians in both samples, and no systematic difference is observed in $R_e/R_{25}$ 
for spectra obtained from different database, instruments and samples. A few CIG are found at higher redshift, but with 
no significant increase on the ratio $R_e/R_{25}$. In general, the similar low means imply that in both samples 
the light from the spectra come from comparable regions in the most central part of the galaxies ($\sim$ 1-2 kpc$^{2}$). 
Therefore, we conclude that there is practically no difference in aperture between the spectra that could influence our results.

\subsection{Morphology distributions} \label{Morph}

It is well know that galaxies in the nearby universe follow a morphology-density relation (Dressler \cite{Dressler80}; 
Whitmore  et al. \cite{Whitmore93}; Deng et al. \cite{Deng08}), according to which early type galaxies are more clustered 
than later type galaxies. The HCG are no exception, the fraction of galaxies with early type morphologies being generally 
higher than in the field (Hickson \cite{Hickson82}; Williams \& Rood \cite{Williams87}; Hickson et al. \cite{Hickson88}). 
Therefore, a statistical comparison of SFH between extreme density regimes needs to account for differences in morphological 
distribution. For our study, we have consequently separated our samples in three distinct morphological classes: galaxies 
in the morphological bin range E-S0 are identified as Early-type galaxies (EtG), galaxies in the morphological bin range 
S0a-Sb are identified as Early-type Spirals (EtS) and the rest, Sbc and later, are identified as Late-type Spirals (LtS). 
In Table~\ref{gals}, we present the statistics for the distribution of morphological classes in our two samples. It can 
clearly be seen that in the HCG sample, the fraction of early types galaxies (both EtG and EtS) is significantly higher 
than in the CIG sample.

\begin{table}
\caption{Morphological class distributions in the samples}
\label{gals}      
\centering                                      
\begin{tabular}{llcc}         
\hline\hline                       
Morphological class  & Types &  HCG  & CIG  \\
\hline
No. of gal.          &  & 210    & 309  \\ 
Early-type galaxies (EtG) & E-S0   & 38\%   & 16\%  \\
Early-spirals       (EtS) & S0a-Sb & 37\%   & 28\%  \\
Late-spirals        (LtS) & Sbc and later    & 25\%   & 56\%  \\
\hline 
\end{tabular}
\end{table}

\subsection{Luminosity and mass distributions} \label{Lum}

To compare the physical characteristics of the galaxies in our two samples we have determined the luminosities using 
the total apparent corrected B-magnitudes compiled in HYPERCAT (Paturel et al. \cite{Paturel94,Paturel97}). These 
magnitudes were corrected for galactic and internal extinction, and k-correction. Such information is missing only for 
HCG 31q. Also, for most of the galaxies in both samples, that is, 80\% and 96\% of the total HCG and CIG respectively, 
we have extracted the $K$-band magnitudes at the 20 mag arcsec$^{-2}$ isophote, as compiled in 
{\it 2MASS}\footnote{http://irsa.ipac.caltech.edu} extended source catalog (Jarrett et al. \cite{Jarrett00}; 
Skrutskie et al. \cite{Skrutskie06}). These magnitudes were corrected for galactic extinction (Schlegel et al. 
\cite{Schlegel98}), and k-correction following Kochanek et al. (\cite{Kochanek01}). To deduce the masses of the galaxies, 
we have adopted a solar value of M$_{\odot,K}=3.28$ and used the mass luminosity ratios, M/L, for different morphological 
types as determined in the K band by James et al. (\cite{James08}).

\begin{figure}
\centering
\resizebox{0.7\hsize}{!}{\includegraphics{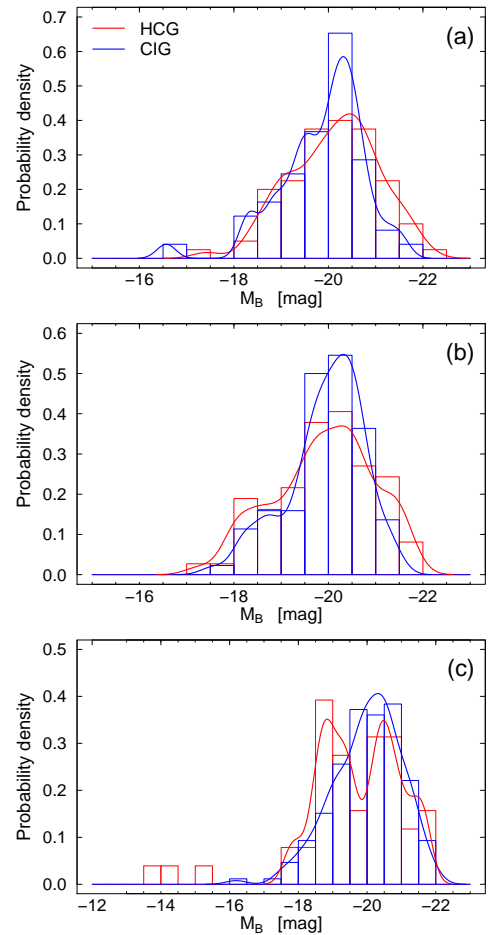}}
\caption{Comparison of the B-band luminosity distribution in our samples: 
a) Early-type galaxies (EtG); b) Early-spirals (EtS); and c) Late-spirals (LtS). 
The continuous curves corresponds to the Kernel density estimation of the probability density of each distribution.
A Gaussian kernel is used.}
\label{histB}
\end{figure}

\begin{figure}
\centering
\resizebox{0.7\hsize}{!}{\includegraphics{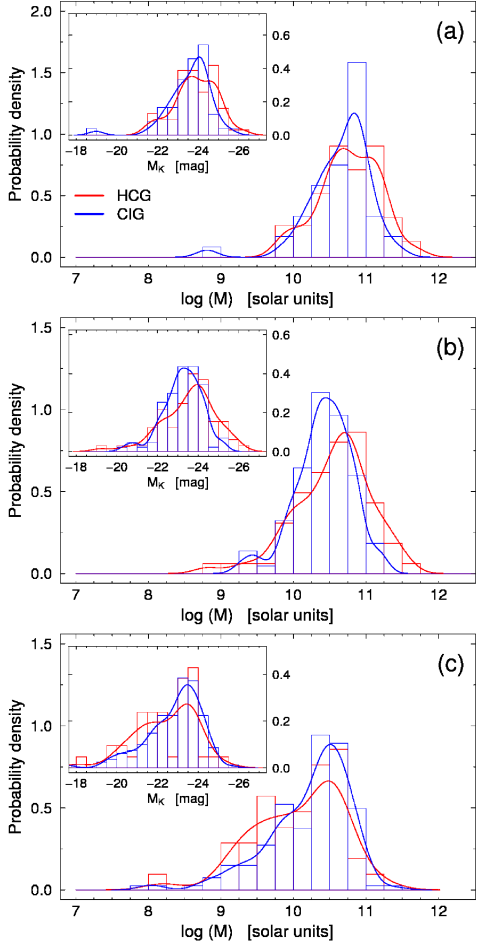}}
\caption{Comparison of the stellar masses and $K$-band luminosity ({\it upper left corners in each panel}) 
distributions in our samples: a) Early-type galaxies (EtG); b) Early-spirals (EtS); and c) Late-spirals (LtS). 
The continuous curves corresponds to the Kernel density estimation of the probability density of each distribution.
A Gaussian kernel is used.}
\label{histK}
\end{figure}

\begin{table}
\caption{Luminosities in B band.}
\label{magsB}      
\centering     
\begin{tabular}{cccccc}         
\hline\hline                       
Morph. class   &   \multicolumn{2}{c}{  M$_B$}  & $\Delta$M$_B  \pm \sigma$ &  $P$ & $S^{a}$  \\
 &   \multicolumn{2}{c}{(mag)}  &  (mag)                      &   &    \\
           &                 HCG & CIG     &                           &      &     \\
\hline
EtG            & $-20.1$  & $-19.9$ & $0.25\pm0.17$ & $0.14$ & $1.5\sigma$  \\
EtS            & $-19.9$  & $-19.9$ & $0.05\pm0.15$ & $0.75$ & ...  \\
LtS            & $-19.9$  & $-20.0$ & $0.13\pm0.18$ & $0.46$ & ...  \\
\hline\hline 
\end{tabular}
\begin{list}{}{}
\item[$^{\mathrm{a}}$] Values $\le 1\sigma$ are not shown.
\end{list}
\end{table}

\begin{table*}
\caption{Luminosities in K band and corresponding masses.}
\label{magsK}      
\centering                                      
\begin{tabular}{ccccccccc}         
\hline\hline 
Morph. class   &   \multicolumn{2}{c}{M$_K$}  & $\Delta$M$_K  \pm \sigma$   &  \multicolumn{2}{c}{log($M/M_{\sun}$) }   
& $\Delta$log($M/M_{\sun}) \pm \sigma$ &  $P$ & $S^{a}$  \\
                       &   \multicolumn{2}{c}{(mag)}  &  (mag)                      &  &  &   &    \\  
                &   HCG  &    CIG              &                             &    HCG     &          CIG          &                 &   &    \\
\hline
EtG       & $-23.9$ & $-23.6$  & $0.30\pm0.21$ & $10.8$ & $10.6$ & $0.12\pm0.08$  & $0.15$ & $1.5\sigma$ \\
EtS       & $-23.5$ & $-23.5$  & $0.30\pm0.20$ & $10.5$ & $10.4$ & $0.12\pm0.08$  & $0.13$ & $1.5\sigma$ \\ 
LtS       & $-22.5$ & $-22.8$  & $0.26\pm0.24$ & $10.1$ & $10.2$ & $0.11\pm0.10$  & $0.27$ & $...$ \\   
\hline\hline 
\end{tabular}
\begin{list}{}{}
\item[$^{\mathrm{a}}$] Values $\le 1\sigma$ are not shown.
\end{list}
\end{table*}

In Fig.~\ref{histB} we compare the HCG and CIG luminosity distributions in B as found in the three morphological classes. 
In the HCG a small fraction of EtG are more luminous than in the CIG. The EtS on the other hand have similar B 
luminosity distributions in both samples, with the HCG showing a slightly larger dispersion than the CIG. The LtS in 
the HCG show a bimodal distribution which is not observed in the CIG. This is in agreement with the study of disk galaxies 
in HCG performed by Iglesias-P\'aramo \& V\'ilchez (\cite{Iglesias98}). 

In Fig.~\ref{histK} we show the distributions for the masses. The distributions for the K luminosities are shown as insets 
in the upper left of the graphics. From these distributions we observe a larger fraction of EtG and EtS in the HCG with 
higher luminosities, consistent with these galaxies are more massive than in the CIG sample. 

In Table~\ref{magsB} we present the mean absolute magnitudes of the galaxies in the HCG and CIG samples as measured in 
the different morphological classes. In the last three columns we report the differences in mean between the two samples, 
the probability, $P$, that the difference observed is due to chance in the selection of the samples, and the statistical 
significance, $S$, of the difference. These values were determined by applying 
a new parametric ANOVA model introduced by Hothorn et al. (\cite{Hothorn08}) and developed for the R software by Herberich 
et al. (\cite{Herberich10})\footnote{http://www.r-project.org}. The max-{\it t} test takes into account possible non-normal 
distribution, unequal variances and unbalanced sample sizes. The statistical test confirms that the EtG in the HCG are 
marginally more luminous, and the EtS and LtS show no significant differences.   

In Table~\ref{magsK} we present the mean absolute magnitudes in K and the corresponding mean masses for the HCG and CIG 
galaxies as measured in the different morphological classes. The results of the max-{\it t} test confirm that the EtG 
and EtS in the HCG are marginally more luminous in K and consequently slightly more massive than in the CIG sample. No 
significant difference is observed for the LtS galaxies. 

In Fig.~\ref{histB} the low luminosity tail at $M_B\ge -16$ is produced by only three galaxies: HCG 54b, c, and d. Since 
we do not observed galaxies as weak as these in the rest of our HCG sample or in the CIG sample, we have discarded them 
from our subsequent analyses. Note that according to Verdes-Montenegro et al. (\cite{Verdes02}) HCG 54 could be the 
remnant of the merger of two galaxies at an advanced stage of evolution, not a CG.

\section{Spectral synthesis analysis method} \label{method}

To determine the SFH we used the stellar population synthesis code \begin{scriptsize}STARLIGHT\end{scriptsize} 
(Cid Fernandes et al. \cite{Cid05}; Mateus et al. \cite{Mateus06}; Asari et al. \cite{Asari07}). Before running 
\begin{scriptsize}STARLIGHT\end{scriptsize}, all the spectra in our samples were first corrected for foreground 
Galactic extinction using the dust maps determined by Schlegel et al. (\cite{Schlegel98}), and then shifted to 
their respective rest frame. The spectra were also resampled with a spectral step $\Delta \lambda$ = 1~\AA, within 
the range 3600 to 7200~\AA, and normalized to the median flux of the continuum within the wavelength window 4530 
to 4580~\AA. \begin{scriptsize}STARLIGHT\end{scriptsize} was then run, using a combination of N$_\star$ Simple 
Stellar Populations (SSP) from the evolutionary synthesis models of Bruzual \& Charlot (\cite{Bruzual03}). These 
models were build based on the medium-resolution Isaac Newton Telescope library of empirical spectra (MILES; 
S{\'a}nchez-Bl{\'a}zquez et al. \cite{Sanchez06}), following the stellar evolutionary tracks of Padova 1994 (Alongi 
et al. \cite{Alongi93}; Bressan et al. \cite{Bressan93}; Fagotto et al. \cite{Fagotto94}), where the initial mass 
function (IMF) of Chabrier (\cite{Chabrier03}), and the extinction law of Cardelli et al. (\cite{Cardelli89}) were 
applied. Our selected base is composed of $N_\star=$150 SSP, and includes 25 different stellar population ages 
in the range $1$~Myr to $18$~Gyr, each with six metallicities. For each galaxy, a mask file was created to exclude 
spectral ranges around important emission lines ([OII], [NeIII], $H_\delta$, $H_\gamma$, $H_\beta$, [OIII], HeI, 
[OI], $H_\alpha$, [NII], [SII]), the ISM absorption feature (NaD) and bad pixels.

As a result of running \begin{scriptsize}STARLIGHT\end{scriptsize} an optimal continuum template spectrum is produced 
for each galaxy, from which we can deduce important information about the stellar populations. In this study we
concentrate on the SSP fractions at different ages. Following Cid Fernandes et al. (\cite{Cid05}), we have derived the
light-weighted average ages, $\langle {\rm log}\; t_{\star}\rangle_{\rm L}$, and the mass-weighted average ages, 
$\langle {\rm log}\; t_{\star}\rangle_{\rm M}$, which are equal respectively to:

\begin{equation} \langle {\rm log}\; t_{\star}\rangle_{\rm L}=\sum^N_{j=1} x_{\rm\,j}\, {\rm log}\; t_{\rm\,j};
\;\;\;\;\;\;\;\; \langle {\rm log}\; t_{\star}\rangle_{\rm M}=\sum^N_{j=1} \mu_{\rm\,j}\, {\rm log}\; t_{\rm\,j},
\end{equation} 

\noindent where $x_{\rm j}\; (j=1,...,N_\star)$  are the fractional contributions to the model flux weighted by light 
of the SSP with ages $t_j$, and $\mu_{\rm j}$ are the mass fractional contributions obtained by using the model 
light-to-mass ratios at the normalization $\lambda_0$. The SFH is obtained by tracing the population vector 
$x_{\rm j}$ or $\mu_{\rm j}$ as a function of $t_j$. The uncertainties on these values are of the order of $\sim
0.03$dex. This was determined empirically by running STARLIGHT 50 times on 9 spectra with different S/N.

By definition, the two weighted average ages are sensitive to different stellar populations: $\langle {\rm log}\;
t_{\star}\rangle_{\rm L}$ is sensitive to the presence of young stellar populations, because of their higher
contribution in light, while $\langle {\rm log}\; t_{\star}\rangle_{\rm M}$ is sensitive to the less luminous and 
older stellar populations, which form the bulk of galaxy mass. Considering that the uncertainties in both average 
ages are of the same order, a quantitative estimate of the Star Formation Time Scale (SFTS) activity in each 
galaxy can be obtained by calculating the difference of mass-weighted and light-weighted average stellar ages as,

\begin{equation} 
\Delta(t_\star)= 10^{\langle {\rm }\;log\; t_{\star}\rangle_{\rm M}} - 10^{\langle {\rm }\;log\; t_{\star}\rangle_{\rm L}}\;\;\;\;\rm (in\; Gyr)
\end{equation} 

For example, a galaxy having experienced constant star formation over its lifetime, a typical late-type spiral galaxy, 
would be expected to have a large $\Delta(t_\star)$, the value growing with time. On the 
other hand, an early-type galaxy, a typical elliptical galaxy for example, where star formation has stopped long ago, 
would be expected to have a small $\Delta(t_\star)$, the value decreasing with time. Note that a very young star forming 
galaxy would also have a small  $\Delta(t_\star)$, but this is because both light and mass are dominated by young stars, 
and the value would be expected to increase with time.  More important, different $\Delta(t_\star)$ for the same morphology 
and mass in different environments could indicate special formation conditions, which is exactly 
what our present study looks for.

\begin{figure}
\centering
\resizebox{0.95\hsize}{!}{\includegraphics{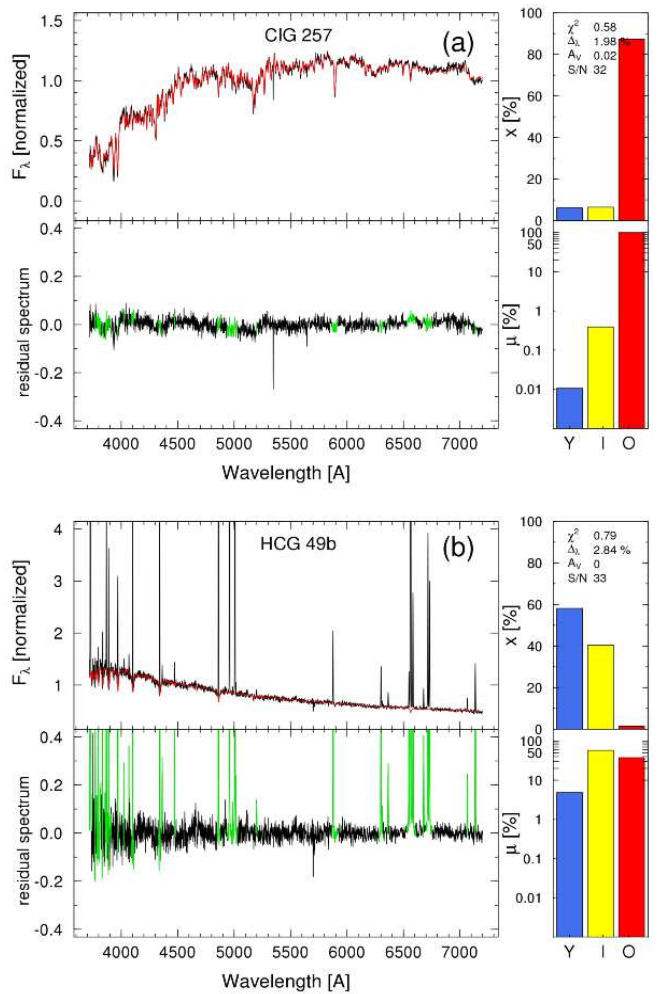}}
\caption{Spectral synthesis results for two galaxies in our samples. Panels are: a) an early-type
galaxy, CIG 257, and b)  a late-type galaxy, HCG 49b. The top-left plot in each panel shows
the synthetic spectrum ({\it red curve}),  plotted over the observed spectrum ({\it black curve})
normalized by the flux at $\lambda =$ 4550$\;\AA$; the bottom-left plot, shows the residual spectrum 
(the difference between the observed and the synthetic continuum spectra), with 
the green regions representing the mask used; the top-right plot shows the percentage contribution of light-weighted 
population vector ($x[\%]$), and the bottom-right plot the percentage contribution of mass-weighted population vector 
($\mu[\%]$ in logarithmic scale), as calculated in each of the three different SSP age groups: young ({\it blue bar}), 
intermediate ({\it yellow bar}), and old stellar population bins ({\it red bar}).}
\label{starlightexam}
\end{figure}

Following Cid Fernandes et al. (\cite{Cid01}, \cite{Cid05}), we have calculated the light ($x_j$) and mass-weighted
($\mu_j$) population vectors in three different age groups: young (Y), with ages in the range $t \le 10^8$ yr (which 
vectors are identified as $X_Y$ and $M_Y$ for light-weight and mass-weight, respectively), intermediate (I), with 
ages within the range $10^8 < t < 5\times10^9$ yr (identified as $X_I$ and $M_I$), and old (O) with stellar population 
ages in the range $t \ge 5\times10^9$ yr  (identified as $X_O$ and $M_O$). For each weighted population vector, we then 
determine the percentage contribution in the three age groups. 

To illustrate our method, we show in Fig.~\ref{starlightexam} the results of applying \begin{scriptsize}STARLIGHT\end{scriptsize} 
on two different galaxies in our samples: an early-type isolated galaxy, CIG 257, and a late-type galaxy in 
CG, HCG 49b. In Fig.~\ref{starlightexam}a it can be seen that in this early-type galaxy the old stellar populations 
dominate both in light and mass (i.e. the largest values are those in $X_O$ and $M_O$). On the other hand, in the late-type 
spiral (Fig.~\ref{starlightexam}b), the young stellar populations dominate the light (i.e. the largest is $X_Y$ and then $X_I$), 
while a mixture of stars of old and intermediate stellar ages dominate the mass (i.e. $M_O$ and $M_I$ are comparable).

\begin{figure}
\centering
\resizebox{1.0\hsize}{!}{\includegraphics[width=8cm]{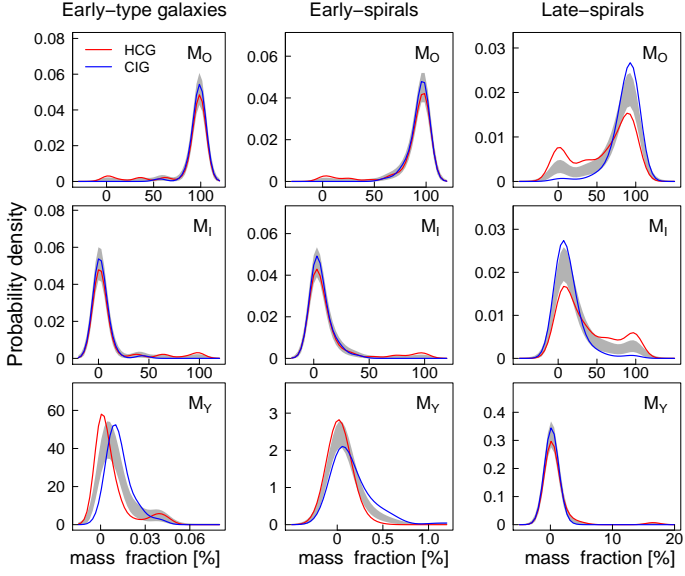}}
\caption{Percentage contribution of mass-weighted population vectors, ({\it top to bottom}) $M_O$, $M_I$, and 
$M_Y$, as measured in the different morphological classes, ({\it left to right)} EtG,  EtS, and LtS galaxies. 
The red and blue curves are the density distributions for the HCG and CIG, respectively. The gray areas are the 
confidence interval models, assuming the two samples comes from the same distribution. In each graph the 
curves are the continuous representation of the distribution. These curves have been done using SM function in   
the R statistical package.}
\label{distrPopsM}
\end{figure}

\begin{figure}
\centering
\resizebox{1.0\hsize}{!}{\includegraphics[width=8cm]{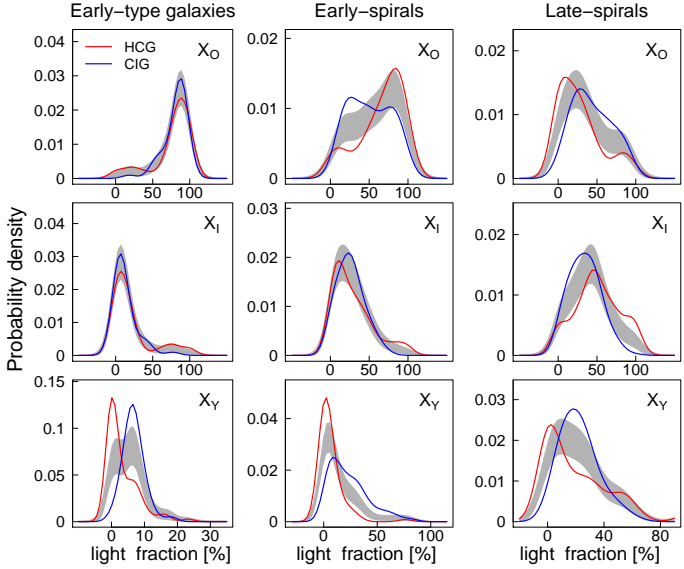}}
\caption{Same as Fig.~\ref{distrPopsM} for the light-weighted population vectors.}
\label{distrPopsL}
\end{figure}

\begin{table}
\caption{Mean values of the mass-weighted population vectors}
\label{tabMfracs}      
\centering                                      
\begin{tabular}{cccccc}         
\hline\hline                       
Population  & \multicolumn{2}{c}{$\mu$}   &  $\Delta \mu \pm \sigma$  & $P $ & $S^{a}$ \\ 
vector      & \multicolumn{2}{c}{$ [\%]$} &     $[\%]$                &      &  \\ 
            &       HCG & CIG             &                           &      &   \\
\hline
\multicolumn{6}{c}{ EtG }\\
\hline
  $M_O$      & $88.4$   & $97.0$  & $8.6\pm3.2$     & $0.007$ &  $2.7\sigma$  \\
  $M_I$      & $11.6$   & $2.9$   & $8.6\pm3.2$     & $0.007$ &  $2.7\sigma$  \\
  $M_Y$      & $0.007$  & $0.013$ & $0.006\pm0.002$ & $0.003$ &  $3.1\sigma$  \\
\hline 
\multicolumn{6}{c}{ EtS }\\
\hline
  $M_O$      & $86.7$   & $93.6$  & $6.9\pm2.9$    & $0.019$ &  $2.4\sigma$  \\
  $M_I$      & $13.3$   & $6.2$   & $7.0\pm2.9$    & $0.017$ &  $2.4\sigma$  \\
  $M_Y$      & $0.06$   & $0.19$  & $0.13\pm0.05$  & $0.008$ &  $2.7\sigma$  \\
\hline 
\multicolumn{6}{c}{ LtS (old)}\\
\hline
  $M_O$      & $89.4$   & $91.3$  & $1.9\pm1.8$    & $0.30$ &  ...  \\
  $M_I$      & $10.4$   & $8.6$   & $1.8\pm1.8$    & $0.30$ &  ...  \\
  $M_Y$      & $0.16$   & $0.16$  & $0.00\pm0.04$  & $0.92$ &  ...  \\
\hline 
\multicolumn{6}{c}{ LtS (young)}\\
\hline
  $M_O$      & $27.8$   & $44.4$  & $16.6\pm8.0$ & $0.04$ &  $2.1\sigma$  \\
  $M_I$      & $62.5$   & $54.8$  & $7.7\pm8.7$  & $0.38$ &  ...  \\
  $M_Y$      & $9.7$    & $0.8$   & $9.0\pm5.8$  & $0.13$ &  ...  \\
\hline 
\hline 
\end{tabular}
\begin{list}{}{}
\item[$^{\mathrm{a}}$] Values $< 2\sigma$ are not shown.
\end{list}
\end{table}

\begin{table}
\caption{Median values of the light-weighted population vectors}
\label{tabXfracs}      
\centering                                      
\begin{tabular}{cccccc}         
\hline\hline                       
Population  & \multicolumn{2}{c}{$x$}   &  $\Delta x \pm \sigma$  & $P $ &  $S^{a}$ \\ 
vector      & \multicolumn{2}{c}{$ [\%]$} &     $[\%]$                &      &  \\ 
            &       HCG & CIG             &                           &      &   \\
\hline
\multicolumn{6}{c}{ EtG }\\
\hline
  $X_O$      & $75.8$   & $80.6$  & $4.9\pm3.9$ & $0.22$           &  $...$  \\
  $X_I$      & $21.0$   & $12.6$  & $8.4\pm4.0$ & $0.04$           &  $2.1\sigma$  \\
  $X_Y$      & $3.2$    & $6.7$   & $3.5\pm0.7$ & $10^{-6}$ &  $4.8\sigma$  \\
\hline 
\multicolumn{6}{c}{ EtS }\\
\hline
  $X_O$      & $65.1$   & $50.8$  & $14.3\pm4.4$  & $0.001$           &  $3.3\sigma$  \\
  $X_I$      & $28.9$   & $25.8$  & $3.2\pm3.5$   & $0.36$            &  $...$  \\
  $X_Y$      & $5.9$    & $23.0$  & $17.5\pm2.4$  & $10^{-11}$ &  $7.4\sigma$  \\
\hline 
\multicolumn{6}{c}{ LtS (old)}\\
\hline
  $X_O$      & $45.9$   & $47.5$  & $1.6\pm5.8$  & $0.79$ &  $...$  \\
  $X_I$      & $34.4$   & $30.5$  & $3.8\pm4.1$  & $0.35$ &  $...$  \\
  $X_Y$      & $20.0$   & $22.0$  & $2.3\pm3.9$  & $0.56$ &  $...$  \\
\hline 
\multicolumn{6}{c}{ LtS (young)}\\
\hline
  $X_O$      & $9.8$   & $9.6$   & $0.3\pm3.8$ & $0.94$ &  $...$  \\
  $X_I$      & $66.3$  & $66.2$  & $0.1\pm7.1$ & $0.99$ &  $...$  \\
  $X_Y$      & $24.0$  & $24.0$  & $0.3\pm7.7$ & $0.97$ &  $...$  \\
\hline\hline 
\end{tabular}
\begin{list}{}{}
\item[$^{\mathrm{a}}$] Values $< 2\sigma$ are not shown.
\end{list}
\end{table}

\section{Results} \label{results}

As explained in Sect.~\ref{Morph} we have separated our two samples in three morphological classes: Early-type 
galaxies (EtG), Early-type spirals (EtS), and Late-type spirals (LtS). In Figs.~\ref{distrPopsM} we show the 
percentage contribution of mass-weighted ($Mw$) population vectors of each age group as found in the different 
morphological classes. In Fig.~\ref{distrPopsL} we show the percentage contribution of the light-weighted 
($Lw$) vectors. In each graph, we have traced the confidence interval model (gray area) constructed by assuming 
the two samples comes from the same distribution. The corresponding mean values of the distributions, the difference 
between means, and the statistical significance given by the statistical tests are reported in 
Tables~\ref{tabMfracs} and \ref{tabXfracs} for the $Mw$ and $Lw$ population vectors, respectively.

The distributions of the $Mw$ and $Lw$ average stellar ages for each morphological class are shown in Figs.~\ref{meanAgeM} 
and \ref{meanAgeL}, respectively. The mean values of the distributions are reported in Table~\ref{tabAge}. As mentioned 
in Sect.~\ref{method}, by studying the difference between the two average ages,  $\Delta(t_\star)$, it is possible to 
quantify the SFTS of the galaxies. Small values of SFTS indicate that both mass and light are produced by the same type 
of stars, which implies that galaxies must have formed all their stars at roughly the same time. The SFTS, therefore, is an 
indicator of how fast a galaxy created its stellar population, or how long the stellar activity was prolonged in this 
galaxy. In Table~\ref{tabScale} we present the mean values of the SFTS and the statistical significance produced by the 
statistical tests comparing these values in the two samples. In the following subsections we discuss the results individually 
for each morphological class. 

\subsection{Early-type galaxies (EtG)} \label{EtGs}

The final number of galaxies in this morphological class amounts to 79 in the HCG and 43 in the CIG sample. We 
have discarded seven EtG, one in the HCG and six in the CIG sample, because we consider them to be atypical for 
their morphology. The nature of the peculiarity observed in these galaxies is discussed briefly in Sect.~\ref{PecEtG}. 
Here, we present the results of what we consider are the typical EtG in both samples.

As can be seen in the left panels of Fig.~\ref{distrPopsM} and \ref{distrPopsL} for the $Mw$ and $Lw$ vectors, 
respectively, old stellar populations dominate the mass ($M_O\sim90$\%) and light ($X_O\sim80$\%) in EtG galaxies 
in both samples. In the HCG there is a small fraction of galaxies showing low $M_O$  and high $M_I$ values. This 
explains the differences observed in Tables~\ref{tabMfracs} and \ref{tabXfracs}. However, the difference in $X_O$ 
vector is not statistically significant. The most important difference observed between the stellar 
population vectors in the two samples is the lower contribution of young stellar populations in the HCG. This difference 
is statistically significant for both, the $Mw$ and $Lw$ vectors.  

\begin{figure*}
\centering
\resizebox{0.9\hsize}{!}{\includegraphics{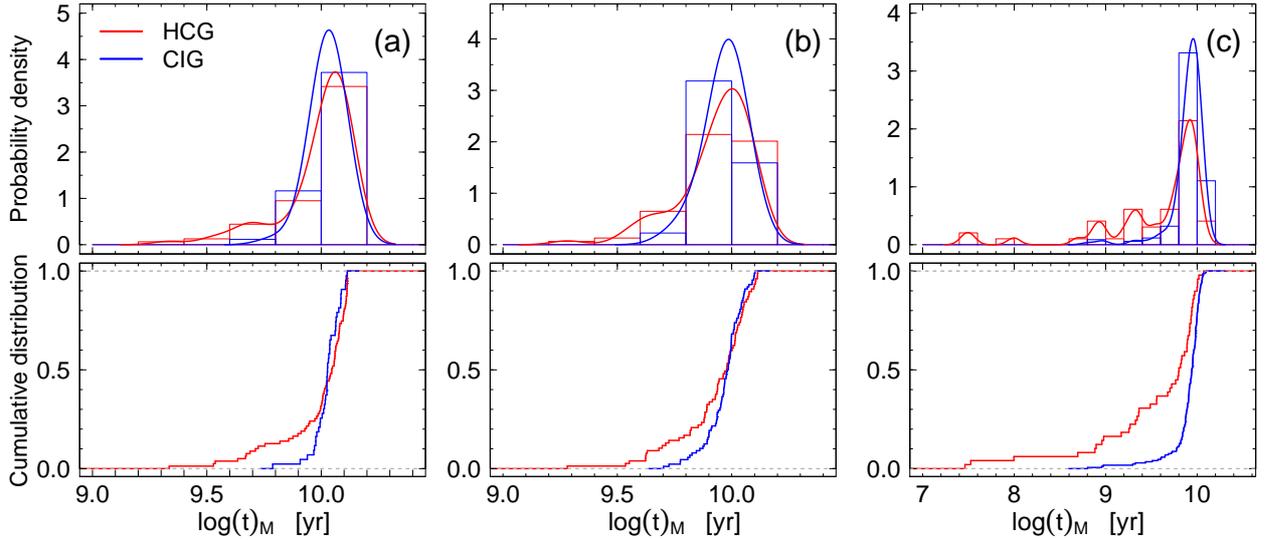}}
\caption{Histogram and cumulative distribution of mass-weighted average stellar ages in the three morphological classes: 
a) EtG, b) EtS, and c) LtS galaxies.}
\label{meanAgeM}
\end{figure*}

\begin{figure*}
\centering
\resizebox{0.9\hsize}{!}{\includegraphics{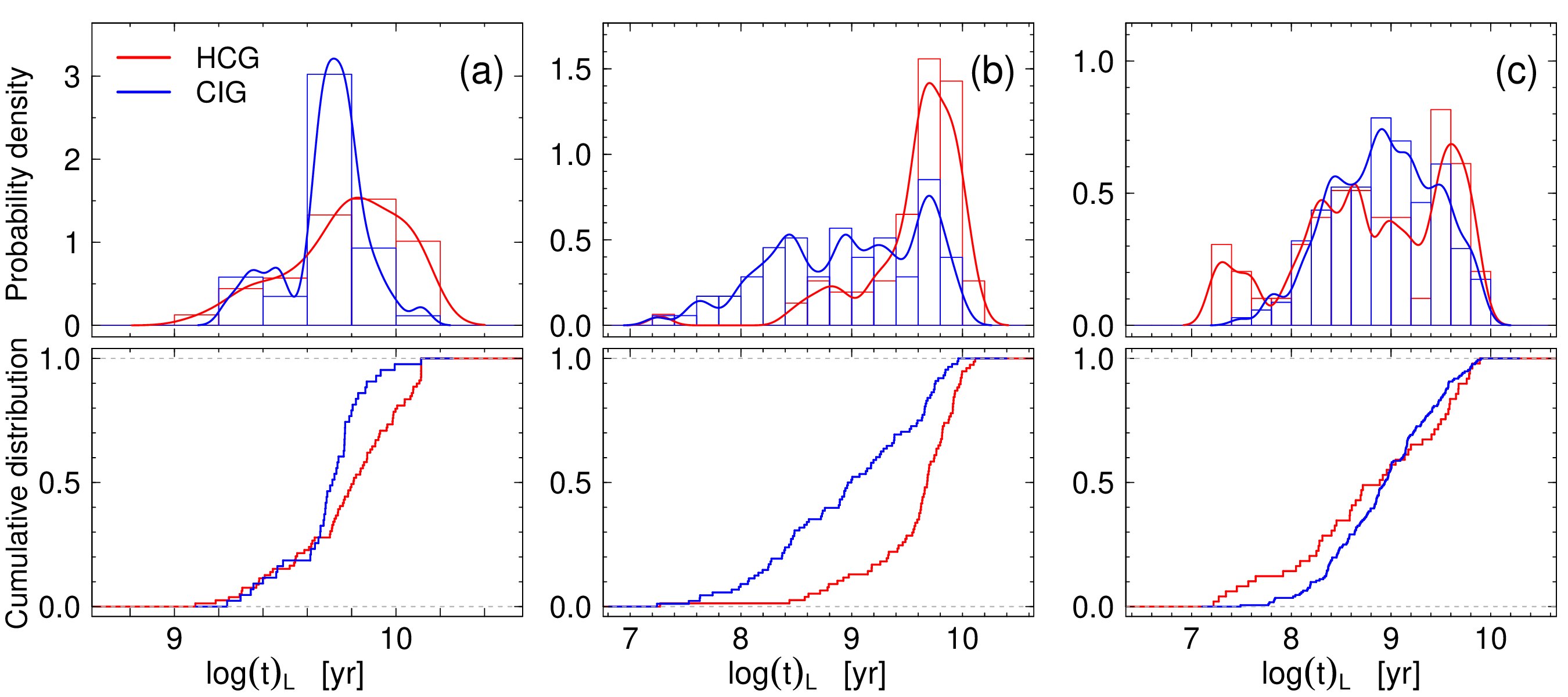}}
\caption{Histogram and cumulative distribution of light-weighted average stellar ages in the three morphological classes: 
a) EtG, b) EtS, and c) LtS galaxies.}
\label{meanAgeL}
\end{figure*}

In Figs.~\ref{meanAgeM}a and \ref{meanAgeL}a  we show the distributions for the $Mw$ and $Lw$ average stellar ages, 
respectively. The mean values are reported in Table~\ref{tabAge}. The EtG in the HCG galaxies are found to be slightly 
older than in the CIG, when the $Lw$ vectors are considered ($\langle {\rm}\; t_{\star}\rangle_{\rm L}$), while no 
difference is detected when the $Mw$ vectors are considered ($\langle {\rm}\; t_{\star}\rangle_{\rm M}$).

\begin{table*}
\caption{Light-weighted  and mass-weighted average stellar ages}
\label{tabAge}
\centering
\begin{tabular}{lcccccccccccc}
\hline\hline
Morphological & \multicolumn{2}{c}{Number}
& \multicolumn{2}{c}{ $\langle {\rm} t_{\star}\rangle_{\rm L}^{a}$} & $\Delta \langle {\rm} t_{\star}\rangle_{\rm L} \pm \sigma$ &  $P$ & $S^{c}$
& \multicolumn{2}{c}{ $\langle {\rm} t_{\star}\rangle_{\rm M}^{b}$} & $\Delta \langle {\rm} t_{\star}\rangle_{\rm M}  \pm \sigma$ &  $P$ & $S^{c}$  \\
 class        &      HCG  & CIG   &         HCG  & CIG      &       &      &    &         HCG  & CIG    &      &      &    \\
\hline
EtG         & 79 & 43  & $ 6.0$  & $ 4.9$ & $1.2\pm0.6$  & $0.04$             & $2.0\sigma$ & $ 9.7$  & $ 10.7$ & $1.1\pm0.6$  & $0.06$ & $...$ \\
EtS         & 77 & 88  & $ 3.7$  & $ 0.9$ & $4.0\pm0.6$  & $2\times10^{-10}$  & $6.8\sigma$ & $ 8.5$  & $ 9.3$  & $1.1\pm0.5$  & $0.04$ & $2.1\sigma$ \\
LtS (old)   & 25 & 155 & $ 0.9$  & $ 0.9$ & $\sim 0.1$   & $0.99$             & $...$       & $ 7.6$  & $ 8.8$  & $1.2\pm0.5$  & $0.02$ & $2.4\sigma$ \\
LtS (young) & 24 & 17  & $ 0.4$  & $ 0.3$ & $0.1\pm0.4$  & $0.73$             & $...$       & $ 1.2$  & $ 3.0$  & $2.4\pm1.1$  & $0.03$ & $2.2\sigma$ \\
\hline\hline
\end{tabular}
\begin{list}{}{}
\item All ages are in units of Gyr.
\item[$^{\mathrm{a,b}}$] These values correspond to the geometric mean of the ages of galaxies in each morphological class (i.e. $10^{\langle {\rm} log\; t_{\star}\rangle_{\rm L}}$ and $10^{\langle {\rm} log\; t_{\star}\rangle_{\rm M}}$).
\item[$^{\mathrm{c}}$] Values $< 2\sigma$ are not shown.
\end{list}
\end{table*}

\begin{table}
\caption{Mean Star Formation activity Time Scales (SFTS)}
\label{tabScale}
\centering
\begin{tabular}{lrrrcr}
\hline\hline
Morphological   &   \multicolumn{2}{c}{ $\langle {\rm} \Delta(t_\star)\rangle$} & $\Delta \pm \sigma$ &  $P$ & $S$  \\
 class          &                 HCG & CIG                         &     
               &      &     \\
\hline
EtG            & $3.3$ & $5.4$ & $2.1\pm0.4$ & $1\times10^{-7}$  &
$5.6\sigma$  \\
EtS            & $3.8$ & $7.3$ & $3.5\pm0.4$ & $4\times10^{-16}$ &
$9.1\sigma$  \\
LtS (old)      & $5.7$ & $7.3$ & $1.6\pm0.6$ & $0.009$           &
$2.6\sigma$  \\
LtS (young)    & $1.2$ & $3.1$ & $1.8\pm0.5$ & $0.001$           &
$3.6\sigma$  \\
\hline\hline
\end{tabular}
\begin{list}{}{}
\item {\bf All ages are in units of Gyr.}
\end{list}
\end{table}

To put the results of our analysis in a more comprehensive form, we trace in Fig.~\ref{ESFHlog} the SFTS, namely 
$\langle {\rm} t_{\star}\rangle_{\rm M}$ vs. $\Delta(t_\star)$. In each graph we also show 
$\langle {\rm} t_{\star}\rangle_{\rm L}$ as a color code bar and we discriminate among different morphological types 
using different symbols. In Fig.~\ref{ESFHlog} we see that most of the EtG in both samples have 
$\langle {\rm} t_{\star}\rangle_{\rm L}$ between $4$ to $9$~Gyr (from green to orange), with a clear tendency for HCG 
to be older than the CIG in this morphological class, this is regardless of the morphological types. 
In particular we find in the HCG an important number of very old galaxies, mostly E-E/S0, but also including some S0, 
all with $\Delta(t_\star)$ remarkably small ($\la 3$~Gyr) and only one EtG in the CIG below this values.  This 
difference is consistent with longer SFTS for the galaxies in the CIG than in the HCG. The statistical tests reported 
in Table~\ref{tabScale} confirm this difference: in the HCG the galaxies have less prolonged SFTS by $\sim$2~Gyr 
than their counterparts in the CIG sample.

The above differences are consistent with slightly different formation processes for the galaxies in the different 
environments. In the HCG the EtG have formed their stars over shorter time scales than similar galaxies in isolation. 
This result is similar to those obtained by Proctor et al. (\cite{Proctor04}), Mendes de Oliveira et al. (\cite{Mendes05}) 
and de la Rosa et al. (\cite{Rosa07}) for elliptical galaxies in CG. This is also a recognized characteristic of 
early-type galaxies forming in clusters of galaxies (Balogh et al. \cite{Balogh99}; Thomas et al. \cite{Thomas05}; 
Bernardi et al. \cite{Bernardi06}; Gobat et al. \cite{Gobat08}; Clemens et al. \cite{Clemens09}; Cooper et al. 
\cite{Cooper10}; and Hoyle et al. \cite{Hoyle11}).

\subsubsection{Nature of peculiar EtG} \label{PecEtG}

\begin{figure}
\centering
\resizebox{1.0\hsize}{!}{\includegraphics[width=8cm]{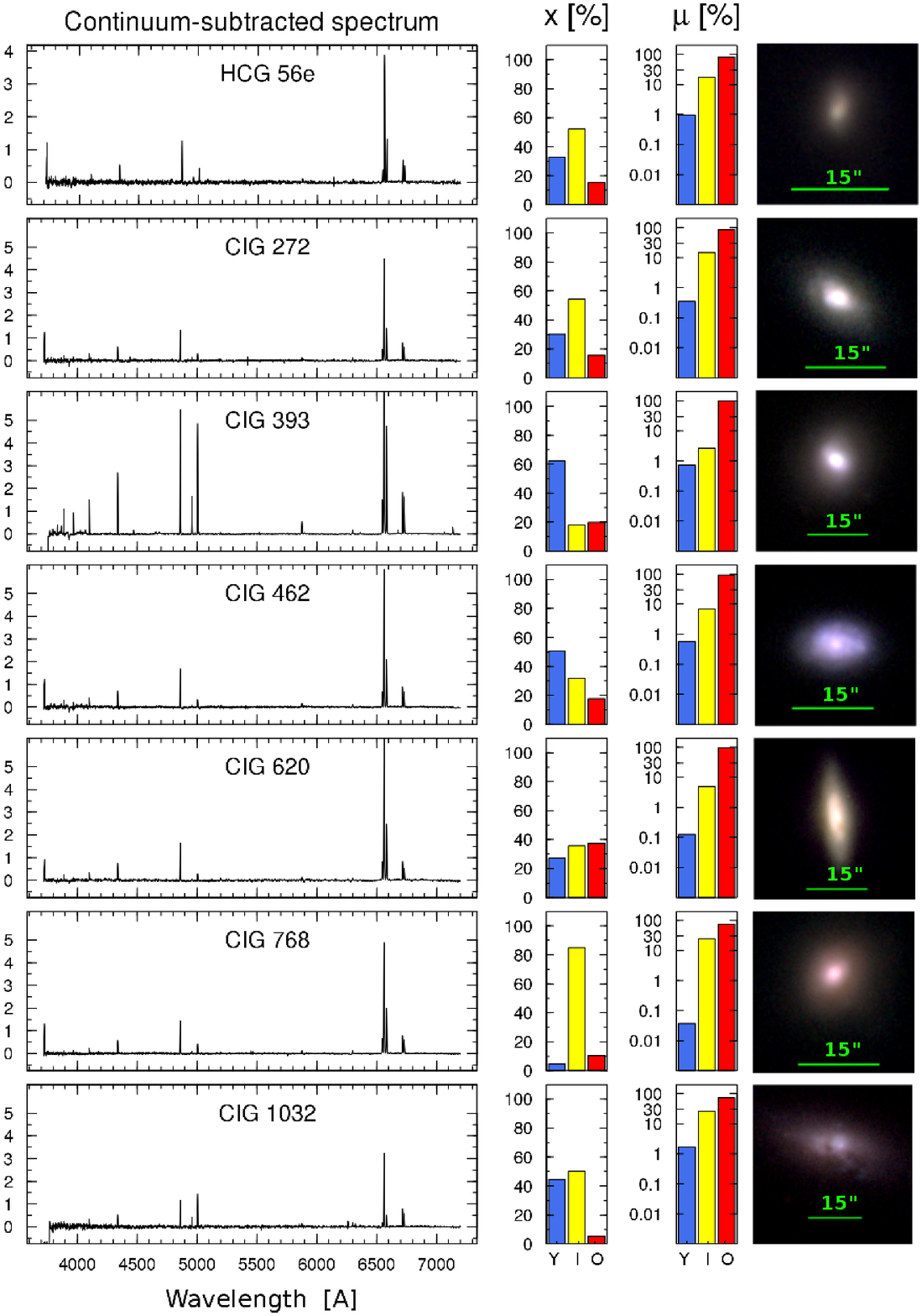}}
\caption{Spectral synthesis results for the seven peculiar EtG. Left panels: continuum-subtracted spectra showing 
the presence of strong emission lines; middle panels: percentage contribution of the $Lw$ and $Mw$ population vectors; 
right panels:{\it gri} composite SDSS images. In these galaxies, the $X_I$ and/or $X_Y$ stellar populations 
tend to dominate the light, which is not typical of EtG.}
 
\label{pecEtG}
\end{figure}

The seven EtG, discarded form the previous analysis due to their atypical spectra for this morphological class, 
are HCG 56e, CIG 272, CIG 393, CIG 462, CIG 620, CIG 768, and CIG 1032. In Figure~\ref{pecEtG} we show the residual 
optical spectra, the population vector analysis and the morphology of these galaxies. In the middle panels of 
Fig.~\ref{pecEtG} we see that although the $M_O$ vectors are dominant, the $X_I$ and $X_Y$ are the highest, suggesting 
intermediate or young stellar populations are producing the light, which is consistent with recent star 
formation episodes. Indeed, the residual spectra, in the left panel of Fig.~\ref{pecEtG}, show intense emission 
lines consistent with a star formation activity (as determined based on a standard diagnostic diagram; Baldwin, Phillips, 
\& Terlevich \cite{Baldwin81}). 

As an independent test, we have also determined the NUV-r color for these galaxies (Yi et al. \cite{Yi05}; Schawinski 
et al. \cite{Schawinski07}), taking the NUV ($\lambda$2270 \AA ) magnitudes from the {\it GALEX} extended source 
catalog (Martin et al. \cite{Martin05}), and the r magnitudes from the SDSS r-band Petrosian magnitudes. The NUV-r 
colors range from 3.1 in CIG 272 to 4.0 in CIG 620 (HCG 56e has 3.2), which is consistent with very young 
($\la 1$~Gyr) stellar populations (Yi et al. \cite{Yi05};  Schawinski et al. \cite{Schawinski07}). 

As for the reason why these EtG show evidence of recent star formation activity, the images in SDSS in 
Figure~\ref{pecEtG} give no obvious clue.  In particular, we detect no trace of interaction. However, according 
to Thomas et al. (\cite{Thomas10}), evidence of recent star formation in EtG is not an uncommon phenomenon. 
These authors also suggested that the fraction of EtG showing current star formation (they call it 
the ``rejuvenation fraction'') in low density environments is 10\% higher than in high density ones, 
which suggests they are generated by internal processes. Taken at face value, this seems to agree with what we 
observe, counting 6 out of 49 (12\%) such galaxies in the CIG compared to only 1 out of 80 in the HCG sample. 

Alternatively, some of these galaxies may have been misclassified morphologically. According to Bitsakis et al. 
(\cite{Bitsakis11}) for example, HCG 56e could be a dusty obscured late-type galaxy. We may see vaguely some spiral 
structures in some of the galaxies in Figure~\ref{pecEtG} (although, not in the optical images of HCG 56e). Dust 
seems also to be present, but this is more a normal characteristic of star forming regions than an indicator of 
a late-type morphology. 

\begin{figure}
\centering
\resizebox{0.9\hsize}{!}{\includegraphics{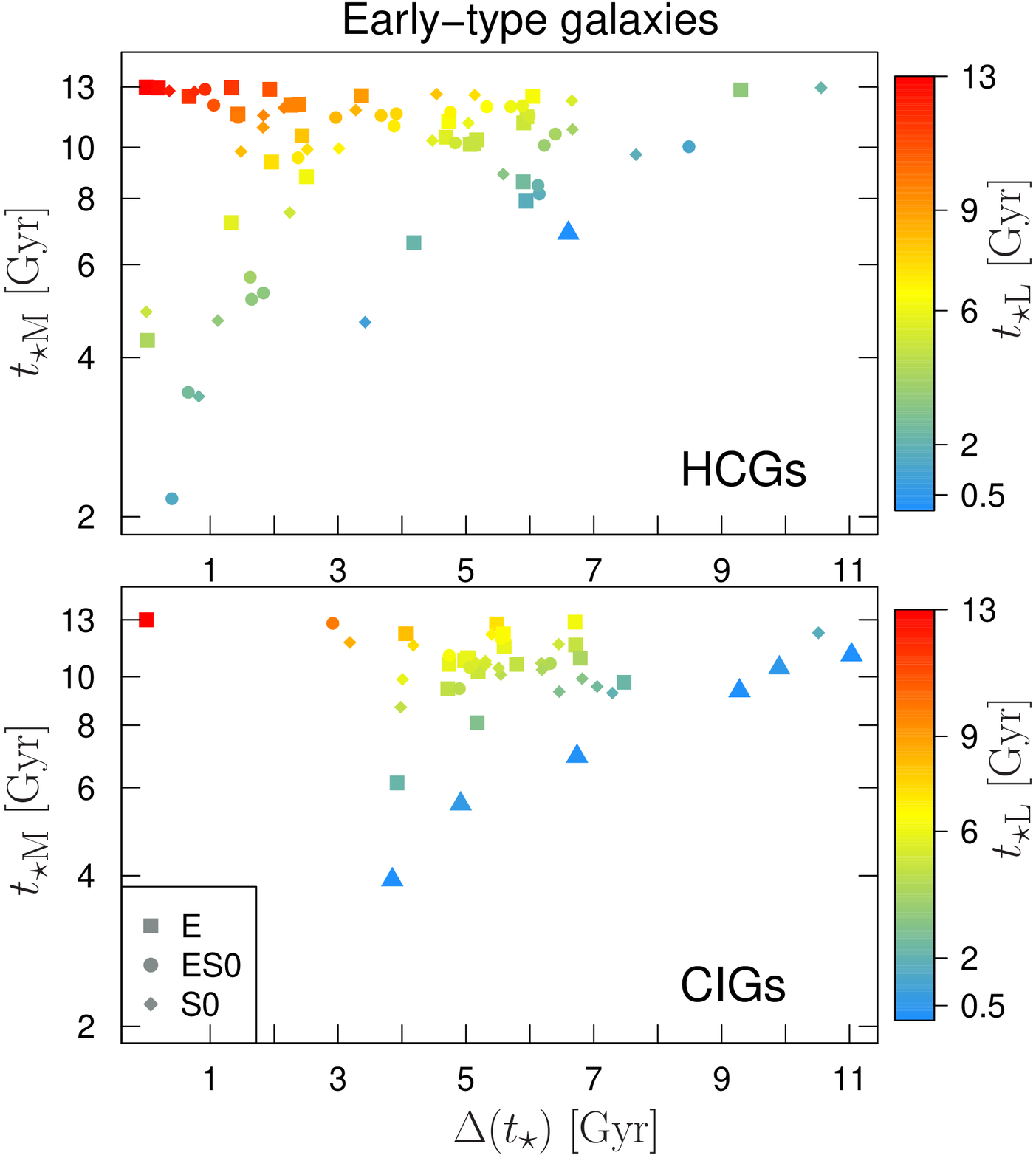}}
\caption{Star Formation activity Time Scale (SFTS) traced in terms of mass-weighted age ${\rm} t_{\star\; \rm M}$ 
as a function of $\Delta(t_\star)$ for EtG in HCG sample ({\it top panel}) and the CIG sample ({\it bottom panel}). 
The color code bar corresponds to light-weighted age ${\rm} t_{\star\; \rm L}$. Triangles represent the peculiar galaxies.}
\label{ESFHlog}
\end{figure}

In Fig.~\ref{ESFHlog}, the peculiar EtG were included as blue triangles. In the CIG sample these galaxies trace 
a sequence where $\langle {\rm}\; t_{\star}\rangle_{\rm M}$ increases almost linearly with $\Delta(t_\star)$, which 
is consistent with constant star formation. The fact that this sequence is similar to the one displayed by the EtS 
and LtS in our sample, as we will show in Sections~\ref{EtS} and \ref{LtS}, reinforced the argument suggesting 
these could be misclassified late type galaxies.

\subsection{Early-type spirals (EtS)} \label{EtS}

We count 77 EtS in the HCG compared to 88 in the CIG samples. Because both samples have a comparable fraction 
of these galaxies (see Table~\ref{gals}) it is clear that, contrary to the EtG, the EtS show no special preference 
to form in dense environments.

In the middle panels of Figs.~\ref{distrPopsM} and \ref{distrPopsL} we show the distributions of $Mw$ and $Lw$ 
stellar population vectors, respectively, for each age group. In Tables~\ref{tabMfracs} and \ref{tabXfracs} we report 
the mean values of these distributions, the difference between means, and the statistical significance determined 
by the statistical tests. In Fig.~\ref{distrPopsM} it can be seen that the distribution of the old and intermediate 
stellar population vectors, $M_O$ and $M_I$,  in both samples are similar. Only a small fraction of galaxies in the 
HCG have lower values of $M_O$ and  higher values of $M_I$, like observed in the EtG.  Again, like for the EtG, the 
distributions for the young stellar population vectors, $M_Y$, are those that differ more drastically comparing 
the two samples. We observe lower values of $M_Y$ and $X_Y$ in the HCG than in the CIG. From the $Lw$ vectors, we 
deduce that the galaxies in the HCG have higher values of $X_O$ and lower values $X_Y$ than in the CIG (no significant 
difference is observe for $X_I$). The EtS in the HCG are therefore dominated by older stellar populations than in 
the CIG and are clearly deficient in young stellar populations.  

Except for the small fraction of galaxies in the HCG with strong contribution from intermediate age populations, which 
produce slightly younger ages in $\langle {\rm}\; t_{\star}\rangle_{\rm M}$, the distributions of $Mw$ average stellar 
ages in Fig~\ref{meanAgeM}b are comparable in the two samples. This result is similar to the one obtained for the EtG. 
Once again, the most remarkable difference appears for the $Lw$ average ages, $\langle {\rm}\; t_{\star}\rangle_{\rm L}$. 
In the HCG, the EtS are older that in the CIG (cf. Fig.~\ref{meanAgeL}b. and Table~\ref{tabAge}). The difference seems 
even higher than what we observed for the EtG, the galaxies in the HCG looking older than in the CIG by as much as 
$\sim 4$~Gyr. Also noted in Fig.~\ref{meanAgeL}b, the dispersion in age which is much narrower in the HCG than in the 
CIG sample. 

\begin{figure}
\centering
\resizebox{0.9\hsize}{!}{\includegraphics{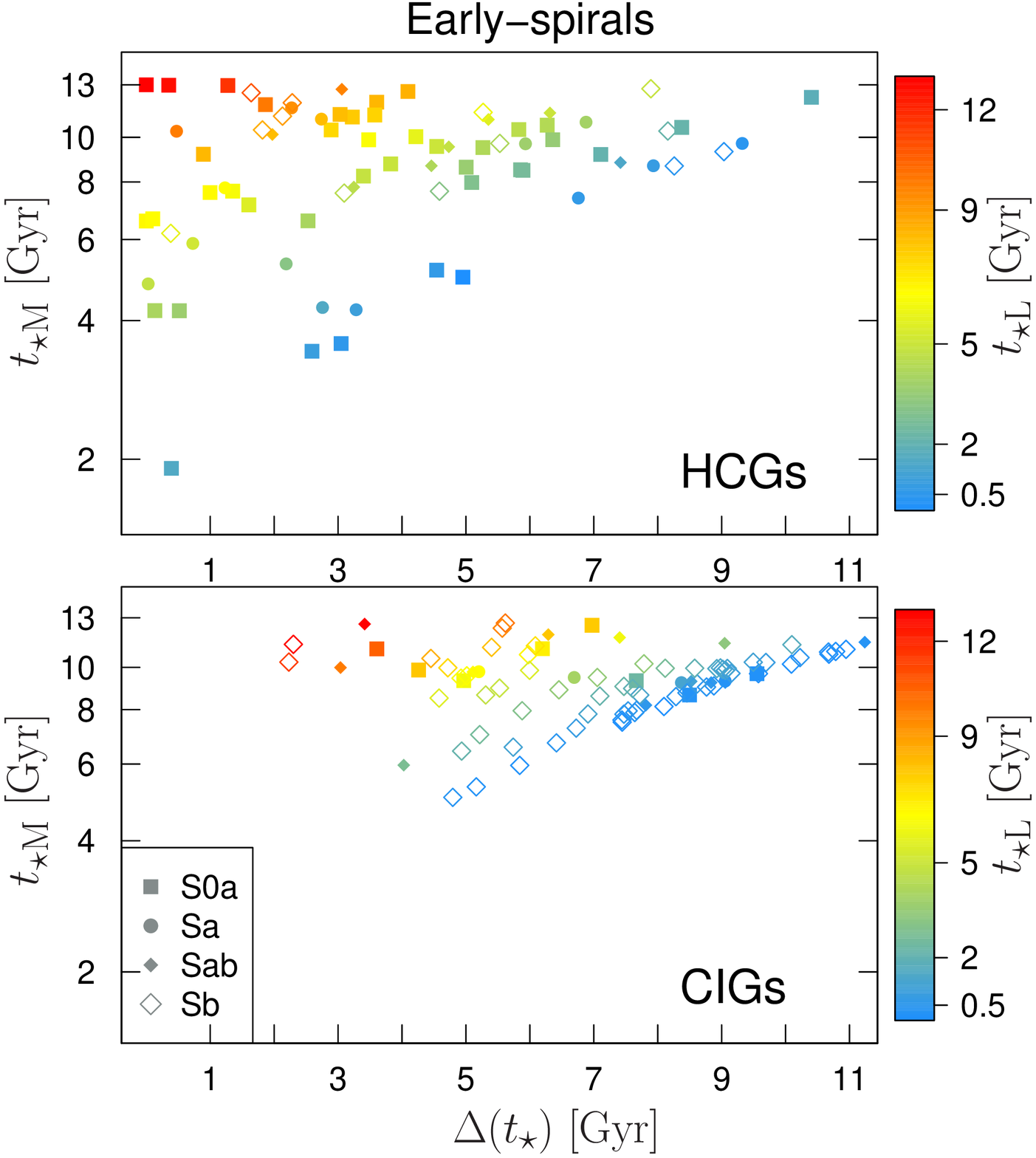}}
\caption{Same as Fig.~\ref{ESFHlog} for EtS galaxies.}
\label{ESSFHlog}
\end{figure}

Once again, these differences become easier to understand in terms of the SFTS. In Fig.~\ref{ESSFHlog} we observe the same 
difference than before for the EtG. In the HCG, the EtS are older than in the CIG and most of them have small values of 
$\Delta(t_\star)$, which is not observed in the CIG sample. The EtS in the HCG have lower SFTS than similar galaxies in 
isolation, indicating that these galaxies in compact groups have less prolongated star formation activity. This is 
confirmed by the statistical tests presented in Table~\ref{tabScale}, where we find on average an SFTS $\sim 3$~Gyr 
shorter in the HCG than in the CIG and we do not distinguish any trend with the morphological type. 

\subsection{Late-type spirals (LtS)} \label{LtS}

In this morphological class we find again a large difference in the number of galaxies between the two samples. We count 
only 49 LtS in the HCG compared to 172 in the CIG sample. Clearly, galaxies in isolation are mostly late-type spirals. 
This is an important difference that must be taken into account. Whatever the processes acting in CG, 
these are related to a change towards earlier-type morphologies.

In the right panels of Figs.~\ref{distrPopsM} and \ref{distrPopsL} we present the percentage contribution of the $Mw$ and 
$Lw$ stellar populations, respectively, for the LtS in our two samples. Contrary than in the CIG, the distribution of $M_O$ 
in the HCG (Fig.~\ref{distrPopsM}) presents a bimodal distribution. A considerable fraction of LtS in the HCG have values 
$M_O < 70$\%, which means that they formed up to 70\% of their stars during the last $5$~Gyr. This low fraction of $M_0$ it 
found in 49\% of the LtS in the HCG galaxies (24 out of 49) compared to only 10\% (17 out of 172) of CIG. To take into account this important difference 
between the two samples, we have separated our samples of LtS galaxies in two subclasses: LtS having $M_O > 70\%$ are classified 
as old, and those that do not meet this criterion are classified as young. In Tables~\ref{tabMfracs} and ~\ref{tabXfracs} 
we present separately the mean values of the $Mw$ and $Lw$ population vectors for old and young LtS. Also, in 
Table~\ref{tabAge} we report for the average stellar ages the mean values of the distributions, the differences between the means, 
and the statistical significance of the differences between the samples.

The presence of the young LtS in the HCG explains the differences observed in Figs.~\ref{distrPopsM} and \ref{distrPopsL}. 
No significant differences are observed in Table~\ref{tabMfracs} (except for a even smaller $M_O$ in the young LtS in HCG than 
in the CIG sample) and Table~\ref{tabXfracs} when the distinction between young and old LtS is taken into account. The presence 
of the young LtS in the HCG also explains the younger $Mw$ average stellar ages in Fig.~\ref{meanAgeM}c, which is confirmed 
in Table~\ref{tabAge}. Most of the young LtS are smaller in size than the old LtS, with a median value in diameter at 25 mag 
arcsec$^{-2}$ isophote of D$_{25}=$15 kpc, both in the HCG and CIG samples (for the old LtS this value is $\sim $24 kpc) and 
have masses of the order log(M/M$_{\sun})<10$. Interestingly, we found that 50\% of the young LtS belong to compact groups 
where most of the members are also LtS. These are HCG 31a,b,c,g,q, HCG 49a,b,c,d, and HCG 100b,c,d.  These galaxies have a 
median value of $\langle {\rm}\; t_{\star}\rangle_{\rm L} \sim$ 200 Myr, confirming they are possible examples of very young 
groups as it has been previously suggested in the literature (Iglesias-P\'aramo \& V\'ilchez \cite{Iglesias97}; V\'ilchez \& 
Iglesias-P{\'a}ramo, J. \cite{Vilchez98}; Verdes-Montenegro et al. \cite{Verdes05}; de Mello et al. \cite{Demello08}; 
Torres-Flores et al. \cite{Torres10}).

\begin{figure}
\centering
\resizebox{0.9\hsize}{!}{\includegraphics{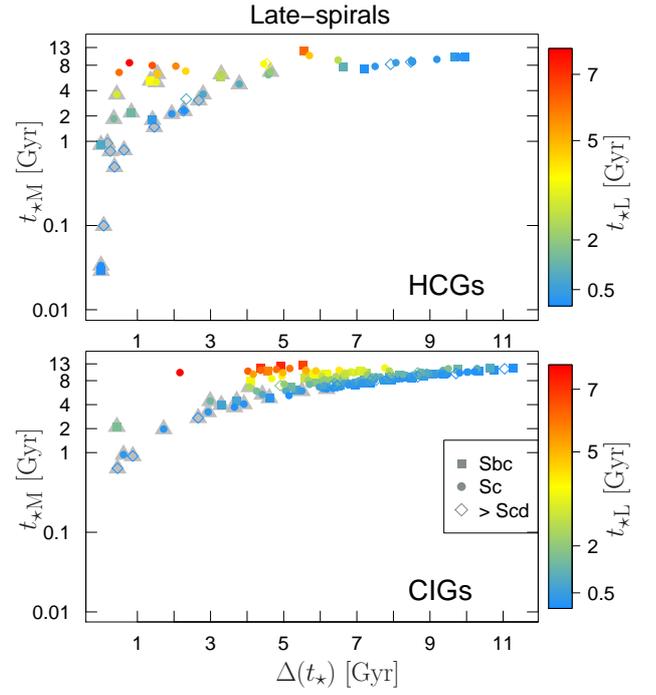}}
\caption{Same as Fig.~\ref{ESFHlog} for LtS galaxies. Young LtS galaxies in the samples are indicated with gray triangles.}
\label{LSSFHlog}
\end{figure}

In Fig.~\ref{LSSFHlog} we show the SFTS for the LtS and report the mean values in Table~\ref{tabScale}. Here we have to 
be careful in interpreting the results. The young LtS in the HCG have small $\Delta(t_{\star})$ (gray triangles in 
Fig.~\ref{LSSFHlog}), mainly because they show young ages with roughly similar values of $\langle {\rm}\; t_{\star}\rangle_{\rm M}$ 
and $\langle {\rm}\; t_{\star}\rangle_{\rm L}$,  which is consistent with very young star forming galaxies. The old LtS on the 
other hand, show the same trend as observed before for the EtG and EtS morphological classes, with a SFTS shorter in the HCG 
than in the CIG by almost $2$~Gyr.

\section{Discussion and conclusions} \label{discussion}

\begin{figure}
\resizebox{0.97\hsize}{!}{\includegraphics{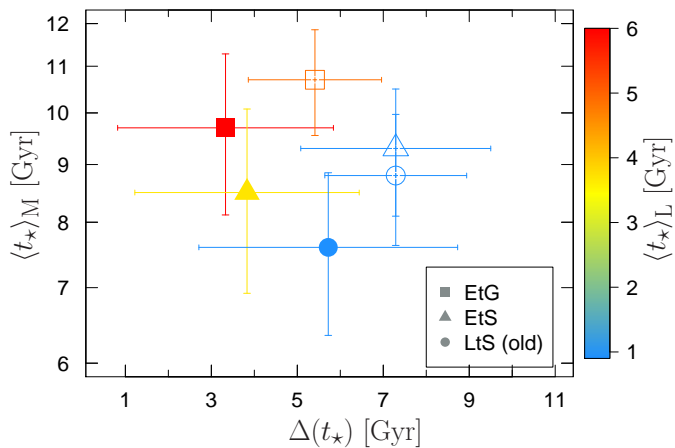}}
\caption{Mean SFTS for EtG, EtS, and old LtS galaxies in HCG ({\it filled symbols}) and CIG samples 
({\it open symbols}). Error bars are $\pm \sigma$} of the mean values.
\label{AllSFHlog}
\end{figure}

To summarize our analysis, we show in Fig.~\ref{AllSFHlog} the mean SFTS as measured in the three morphological classes, 
EtG, EtS and LtS (old subclass only), comparing the HCG  with the CIG. The most remarkable difference observed between the 
galaxies in the HCG (filled symbols) and CIG (open symbols) is the less prolongated star formation activity, namely SFTS, 
in the HCG compared to the CIG sample. The SFTS is shorter by $\sim 2$~Gyr in the EtG and LtS, and even shorter by $\sim 3.5$~Gyr 
in the EtS. These differences are independent of the morphology, mass, and luminosity of the galaxies. 

\begin{figure}
\centering
\resizebox{1.0\hsize}{!}{\includegraphics{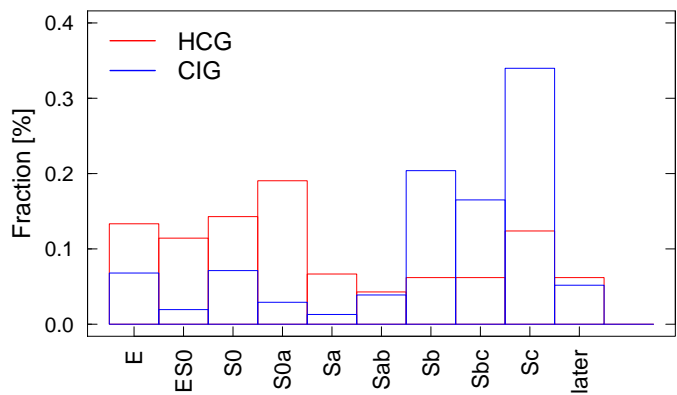}}
\caption{Morphology type distribution of galaxies in the HCG and CIG samples.}
\label{T_DISTR}
\end{figure}

For the EtG, we conclude that they have formed their stars more rapidly. In other words, our results suggest that the EtG in 
the HCG show higher astration rates than in the CIG (i.e., they transformed their gas more efficiently into stars). Higher 
astration rates for the EtG imply more frequent mergers (Coziol et al. \cite{Coziol11}).  This is consistent with the 
hierarchical galaxy formation model, according to which gas rich protogalaxies in dense environments experience more 
frequent mergers, producing galaxies with earlier-type morphologies (i.e., galaxies with bigger bulges). However, since the 
CG are forming from lower mass density fluctuations than clusters of galaxies, very few of these mergers would produce 
massive elliptical galaxies. This picture is consistent with the morphological distribution in the HCG, as seen in 
Fig.~\ref{T_DISTR}, which shows that these systems are dominated by lenticular galaxies.

\begin{figure}
\centering
\resizebox{0.7\hsize}{!}{\includegraphics{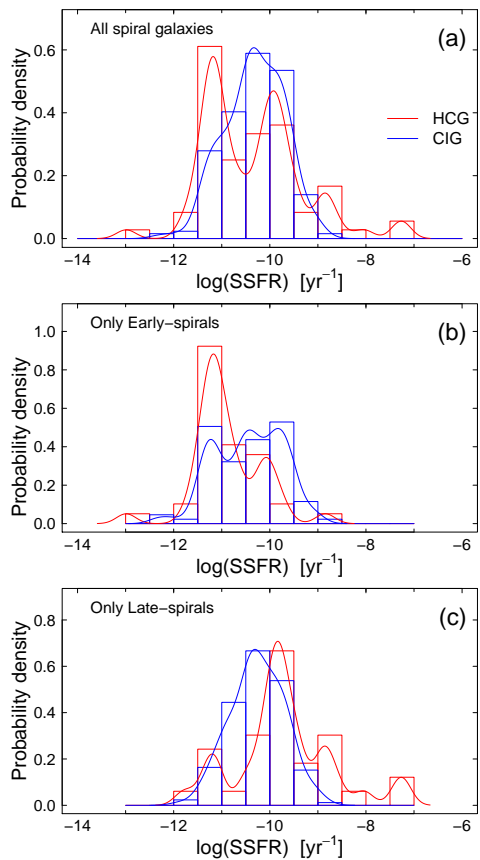}}
\caption{Comparison of the distributions of SSFR in the samples: 
a) both EtS and LtS; b) only EtS; and c) only LtS galaxies.}
\label{SSFRsp}
\end{figure}

For the EtS and LtS, the interpretation is not as obvious as for the EtG. In these galaxies, we cannot eliminate a priori 
tidal gas stripping as a possible mechanism to reduce the SFTS. These would be examples of galaxies that fell into the CG 
later. To test the gas stripping hypothesis, we have calculated the Specific Star Formation Rate (SSFR) of all the spiral 
galaxies in our samples (EtS and LtS galaxies). The SSFR, defined as the actual SFR divided by the stellar mass, reflects 
the mass of stars formed during the last $10^8$ yr. The SSFR is estimated using \begin{scriptsize}STARLIGHT\end{scriptsize} 
by integrating the mass-weighted population vector over the young age group $M_Y$ (see Eq. 6 in Asari et al. \cite{Asari07}). 
In Fig.~\ref{SSFRsp}a we compare the distributions of the SSFR in the HCG and  CIG samples. The distribution of SSFR is 
clearly bimodal (i.e. with a gap) for the HCG, while the distribution of the CIG is continuous. The bimodal distribution of 
the HCG is a by-product of a difference in morphology. As before, we distinguish between early and and late type spirals in 
both samples and find that the EtS in the HCG have a distribution of SSFR that peaks at lower values than for the CIG distribution 
(Figs.~\ref{SSFRsp}b). On the other hand, the SSFR of the LtS in HCG peaks at a higher value than in the CIG sample 
(Fig.~\ref{SSFRsp}c). This is consistent with the hypothesis that late-type spirals, still rich in gas, are recent acquisitions 
in CG. By falling into the group these galaxies increase their star formation and rapidly consume their gas (i.e., higher 
astration rates). Thus, by increasing their astration rates and interacting with other galaxies, these new members will 
transform into earlier-type spirals. Note that this interpretation do not eliminate tidal gas stripping completely, which 
surely may also play a role in lowering the SFTS of these galaxies. However, our test indicates that this is not the main 
mechanism. Recent studies, based on UV and IR data, have also found a bimodality (gap) in the distribution of SSFR in HCG 
galaxies and suggested that the high density environment of CGs has accelerated the evolution of galaxies transforming star 
forming galaxies into quiescent galaxies (Johnson et al. \cite{Johnson07}; Tzanavaris et al. \cite{Tzanavaris10}; Walker et al. 
\cite{Walker10}; Bitsakis et al. \cite{Bitsakis10}, \cite{Bitsakis11}).

According to our analysis, galaxies in the HCG have shorter SFTS than in the CIG because they have experienced higher 
astration rates. This explains why galaxies in HCG are generally deficient in gas (Williams \& Rood \cite{Williams87}; 
Verdes-Montenegro et al. \cite{Verdes01}; Borthakur et al. \cite{Bortha10}). The galaxies evolved more rapidly than in 
the field. This is consistent with the hierarchical galaxy formation model which states that CG are structures 
that formed recently from primordial low fluctuations in mass density. As for the time of the formation of the HCG 
the systematic difference in SFTS suggests this could have happened $2$ to $3.5$~Gyr in the past.  

Our observations reinforce the idea that these are examples of relatively young structures as has been discussed by 
Hickson (\cite{Hickson97} and references therein). However, $\sim 3$~Gyr for the time of formation of these structures 
may still seem too old to explain why these systems have not collapsed into more gravitationally bound objects, like a 
giant blue elliptical galaxy (in fact, dynamically speaking, we should have rather expected a smaller version of a 
cD galaxy, i.e. an elliptical galaxy with an inflated envelope). In the introduction, we have briefly mentioned why according to 
the hierarchical galaxy formation model such a final state for the HCG is not plausible. This is because these structures 
formed from very small density fluctuation regions, where there was not enough mass to transform gas rich protogalaxies 
into giant elliptical or cD galaxies.

Does this mean that the HCG must be in some sort of dynamical equilibrium? This hypothesis was tested by Plauchu-Frayn 
\& Coziol (\cite{Plauchu10a}, \cite{Plauchu10b}), who showed that galaxies in the HCG are not in an equilibrium state, 
but merging without gas. Now, it is worth emphasizing what is the main dynamical consequence of the absence of gas on mergers. 
By definition, gravity is a conservative force, which implies that to form a gravitationally bounded structure, 
a system must loose some of its energy. This is where the presence of free gas plays a key role. The interaction of gas-rich 
protogalaxies is a highly dissipative process. The gas is heated then cooled through radiation, and non isotropic 
gravitational interactions produce gas turbulence inducing star formation and triggering the formation of SMBH at 
the center of galaxies. Therefore, the first phase in the formation of galaxies in CG must be relatively fast, because 
of the high level of dissipation of energy produced by the gas. This is consistent with the short SFTS found in HCG and 
the bimodality in the SSFR distribution. However, further interactions between gas-poor galaxies (i.e., dry mergers) is 
a much less dissipative process, and consequently the dynamical time scales for the merger of these systems may be 
longer than initially assumed (a few Gyr), which would explain why we still observe CG in their present states.

To summarize, our analysis reveals that the galaxies in the HCG are older than in the CIG and that their SFTS are shorter by 
$\sim 3$~Gyr. Considering that these are systematic differences, independent of the morphology, of the mass and  
luminosity of the galaxies, we conclude that this can only imply that the galaxies in the HCG formed their stellar 
population faster than the galaxies in the CIG.  We conclude that the main effect of the environment on galaxies 
in the HCG sample is an increase of astration rate, consistent with the fast mergers of gas-rich protogalaxies. This 
is consistent with the hierarchical galaxy formation model, which states that the HCG are relatively recent structures 
that formed from primordial low mass density fluctuations. Based on the systematic difference in SFTS we propose that 
the HCG most probably formed $\sim 3$~Gyr in the past.  The galaxies in the HCG are not in equilibrium but merging 
without gas, which may explain why they have longer dynamical lifetime.

\begin{acknowledgements} 

We are grateful to M.~A. Mart\'inez for making available her sample of HCG spectra and to J. Perea, 
C. Husillos, and J. Islas-Islas for useful comments on this work. I.~P-F acknowledges the postdoctoral fellowship 
grants 145727 and 170304 from CONACyT Mexico and partial support from the Spanish research project AYA2010-15196. 
A.~d.~O. is partially supported by the Spanish research project AYA2010-22111, Junta de Andaluc\'ia TIC114 and 
the Excellence project P08-TIC-3531. T-P acknowledges PROMEP for support grant 103.5-10-4684 and DAIPGto (0065/11).
We thank the anonymous referee for the valuable and detailed comments that improved this manuscript. 
Finally, we acknowledge the use of the following databases and software:

\begin{itemize}

\item Funding for the Sloan Digital Sky Survey (SDSS) has been provided by the Alfred P.Sloan Foundation, the 
Participating Institutions, the National Aeronautics and Space Administration, the National Science Foundation, 
the US Department of Energy, the Japanese Monbukagakusho and the Max-Planck Society.  The SDSS is managed by 
the Astrophysical Research Consortium for the participating Institutions. The Participating Institutions are the
University of Chicago, Fermilab, the Institute for Advanced Study, the Japan Participation Group, Los Alamos 
National Laboratory, the Max-Planck Institute for Astronomy(MPIA), New Mexico State University, University of 
Pittsburgh, University of Portsmouth, Princeton University, the United States Naval Observatory and the University 
of Washington (http://www.sdss.org). (http://www.sdss.org/). 
\item {\it 2MASS} ({\it Two Micron All Sky Survey}) is a joint project of the University of Massachusetts and 
the Infrared Processing and Analysis Center/California Institute of Technology, funded by the National Aeronautics 
and Space Administration and the National Science Foundation.
\item Hyperleda is an information system for astronomy based on the richest catalogue of homogeneous parameters 
of galaxies for the largest available sample (http://leda.univ-lyon1.fr).
\item {\it NED  (NASA/IPAC Extragalactic Database)} is operated by the Jet Propulsion Laboratory, California 
Institute of Technology, under contract with the National Aeronautics and Space Administration.
\item {\it GALEX} ({\it Galaxy Evolution Explorer}) is a NASA Small Explorer, launched in April 2003 and developed 
in cooperation with the Centre National d'Etudes Spatiales of France and the Korean Ministry of Science and Technology. 
\item R is a free software environment for statistical computing and graphics (http://www.r-project.org/). 
\item TOPCAT is an interactive graphical tool for analysis and manipulation of tabular data (Taylor \cite{Taylor05})\\ 
(http://www.star.bris.ac.uk/~mbt/topcat/).
\end{itemize}
\end{acknowledgements}

\begin{thebibliography}{}

\bibitem[2009]{Abazajian09} {Abazajian, K.~N., Adelman-McCarthy, J.~K., Ag{\"u}eros, M.~A., et al. 2009, \apjs, 182, 543}
\bibitem[1999]{Allam99} {Allam, S.~S., Tucker, D.~L., Lin, H. \& Hashimoto, Y. 1999, \apj, 522, 89L}
\bibitem[1993]{Alongi93} {Alongi, M., Bertelli, G., Bressan, A., et al. 1993, \aap, 97, 851}
\bibitem[2012]{Alonso12} {Alonso, S., Mesa, V., Padilla, N., \& Lambas, D.~G. 2012, \aap, 539, 46}
\bibitem[2007]{Asari07} {Asari, N.~V., Cid Fernandes, R., Stasi{\'n}ska, G., et al. 2007, \mnras, 381, 263}
\bibitem[1981]{Baldwin81} {Baldwin, J.~A., Phillips, M.~M., \& Terlevich, R. 1981, \pasp, 93, 5}
\bibitem[1999]{Balogh99} {Balogh, M.~L., Morris, S.~L., Yee, H.~K.~C., et al. 1999, \apj, 527, 54}
\bibitem[2006]{Bernardi06} {Bernardi, M., Nichol, R.~C., Sheth, R.~K., et al. 2006, \aj, 131, 1288}
\bibitem[2010]{Bitsakis10} {Bitsakis, T., Charmandaris, V., Le Floc'h, E., et al. 2010, \aap, 517, 75}
\bibitem[2011]{Bitsakis11} {Bitsakis, T., Charmandaris, V., da Cunha, E., et al. 2011, \aap, 533, 142}
\bibitem[2010]{Bortha10} {Borthakur, S., Yun, M. S., \& Verdes-Montenegro, L. 2010, \apj, 710, 385}
\bibitem[1993]{Bressan93} {Bressan, A., Fagotto, F., Bertelli, G., \& Chiosi, C. 1993, \aap, 100, 647}
\bibitem[2003]{Bruzual03} {Bruzual, G. \& Charlot, S. 2003, \mnras, 344, 1000}
\bibitem[1990]{Byrd90} {Byrd, G. \& Valtonen, M. 1990, \apj, 350, 89}
\bibitem[1994]{Caon94} {Caon, N., Capaccioli, M., D'Onofrio, M., \& Longo, G. 1994, \aap, 286, 39}
\bibitem[1989]{Cardelli89} {Cardelli, J.~A., Clayton, G.~C., \& Mathis, J.~S. 1989, \apj, 345, 245}
\bibitem[2003]{Chabrier03} {Chabrier, G. 2003, \pasp, 115, 763}
\bibitem[2001]{Cid01} {Cid Fernandes, R., Sodr{\'e}, L., Schmitt, H.~R., \& Le{\~a}o, J.~R.~S. 2001, \mnras, 325, 60}
\bibitem[2005]{Cid05} {Cid Fernandes, R., Mateus, A., Sodr{\'e}, L., Stasi{\'n}ska, G., \& Gomes, J.~M. 2005, \mnras, 358, 363}
\bibitem[2009]{Clemens09} {Clemens, M.~S., Bressan, A., Nikolic, B., \& Rampazzo, R. 2009, \mnras, 392, 35}
\bibitem[1994]{Cole94} {Cole, S., Aragon-Salamanca, A., Frenk, C. S., Navarro, J.~F. \& Zepf, S. E. 1994, \mnras, 271, 781}
\bibitem[2010]{Cooper10} {Cooper, M.~C., Gallazzi, A., Newman, J.~A., \& Yan, R. 2010, \mnras, 402, 1942}
\bibitem[1998]{Coziol98} {Coziol, R., Ribeiro, A.~L.~B., de Carvalho, R.~R., \& Capelato, H.~V. 1998, \apj, 493, 563}
\bibitem[2000]{Coziol00} {Coziol, R., Iovino, A., \& de Carvalho, R.~R. 2000, \aj, 120, 47}
\bibitem[2004]{Coziol04} {Coziol, R., Brinks, E., \& Bravo-Alfaro, H. 2004, \aj, 128, 68}
\bibitem[2007]{Coziol07} {Coziol, R. \& Plauchu-Frayn, I. 2007, \aj, 133, 2630}
\bibitem[2009]{Coziol09} {Coziol, R., Andernach, H., Caretta, C. A., Alamo-Mart\'{\i}nez, K.~A. \& Tago, E. 2009, \aj, 137, 4795}
\bibitem[2011]{Coziol11} {Coziol, R, Torres-Papaqui, J.~P., Plauchu-Frayn, I., Islas-Islas, J.~M., et al. 2011, \rmxaa, 47, 361}
\bibitem[2007]{Rosa07} {de la Rosa, I.~G., de Carvalho, R.~R., Vazdekis, A., \& Barbuy, B. 2007, \aj, 133, 330}
\bibitem[2008]{Demello08} {de Mello, D.~F., Torres-Flores, S. \& Mendes de Oliveira, C. 2008, \apj, 135, 319}
\bibitem[2008]{Deng08} {Deng, X.-F., He, J.-Z., \& Wu, P. 2008, \aap, 484, 355}
\bibitem[1980]{Dressler80} {Dressler, A. 1980, ApJS, 42, 565}
\bibitem[2008]{Durbala08} {Durbala, A., del Olmo, A., Yun, M.~S., et al. 2008, \aj, 135, 130}
\bibitem[1994]{Fagotto94} {Fagotto, F., Bressan, A., Bertelli, G., \& Chiosi, C. 1994, \aaps, 104, 365}
\bibitem[1998]{Fujita98} {Fujita, Y. 1998, \apj, 509, 587}
\bibitem[2008]{Gobat08} {Gobat, R., Rosati, P., Strazzullo, V., et al. 2008, \aap, 488, 853}
\bibitem[1996]{Henriksen96} {Henriksen, M. \& Byrd, G. 1996, \apj, 459, 82}
\bibitem[2010]{Herberich10} {Herberich, E., Sikorski, J., \& Hothorn, T. 2010, PLoS, 5, 3}
\bibitem[1982]{Hickson82} {Hickson, P. 1982, \apj, 255, 382}
\bibitem[1988]{Hickson88} {Hickson, P., Kindl, E., \& Huchra, J.~P. 1988, \apj, 331, 64}
\bibitem[1992]{Hickson92} {Hickson, P., Mendes de Oliveira, C., Huchra, J.~P., \& Palumbo, G.~G. 1992, \apj, 399, 353}
\bibitem[1997]{Hickson97} {Hickson, P. 1997, \araa, 35, 357}
\bibitem[2008]{Hothorn08} {Hothorn, T., Bretz, F., \& Westfall, P. 2008, BiomJ, 50, 346}
\bibitem[2011]{Hoyle11} {Hoyle, B., Jimenez, R., \& Verde, L. 2011, \mnras, 415, 2818}
\bibitem[1997]{Iglesias97} {Iglesias-P{\'a}ramo, J. \& V{\'{\i}}lchez, J.~M. 1997, \apj, 479, 1901}
\bibitem[1998]{Iglesias98} {Iglesias-P{\'a}ramo, J. \& V{\'{\i}}lchez, J.~M. 1998, \apj, 518, 94}
\bibitem[1999]{Iglesias99} {Iglesias-P{\'a}ramo, J. \& V{\'{\i}}lchez, J.~M. 1999, \apj, 518, 94}
\bibitem[2008]{James08} {James, P.~A., Prescott, M., \& Baldry, I.~K. 2008, \aap, 484, 703}
\bibitem[2000]{Jarrett00} {Jarrett, T.~H., Chester, T., Cutri, R., et al. 2000, \aj, 119, 2498}
\bibitem[2007]{Johnson07} {Johnson, K.~E., Hibbard, J.~E., Gallagher, S.~C., et al. 2007, \aj, 134, 1522}
\bibitem[1973]{Karachentseva73} {Karachentseva, V.~E. 1973, Soobshch. Spets. Astroﬁz. Obs., 8, 3}
\bibitem[1993]{Kauffmann93} {Kauffmann, G., White, S.~D.~M., \& Guiderdoni, B. 1993, \mnras, 264, 201}
\bibitem[2003]{Kauffmann03} {Kauffmann, G., Heckman, T.~M., Tremonti, C. et al. 2003, \mnras, 346, 1055}
\bibitem[2001]{Kewley01} {Kewley, L.~J., Dopita, M.~A., Sutherland, R.~S., Heisler, C.~A., \& Trevena, J. 2001, \apj, 556, 121}
\bibitem[2001]{Kochanek01} {Kochanek, C.~S., Pahre, M.~A., Falco, E.~E., et al. 2001, \apj, 560, 566}
\bibitem[2005]{Martin05} {Martin, D.~C., Fanson, J., Schiminovich, D., et al. 2005, \apj, 619, 1L}
\bibitem[2008]{Martinez08} {Mart{\'{\i}}nez, M.~A., del Olmo, A., Coziol, R., \& Focardi, P. 2008, \apj, 678, 9}
\bibitem[2010]{Martinez10} {Mart{\'{\i}}nez, M.~A., Del Olmo, A., Coziol, R., \& Perea, J. 2010, \aj, 139, 1199}
\bibitem[2006]{Mateus06} {Mateus, A., Sodr{\'e}, L., Cid Fernandes, R., et al. 2006, \mnras, 370, 721}
\bibitem[1994]{Mendes94} {Mendes de Oliveira, C. \& Hickson, P. 1994, \apj, 427, 684}
\bibitem[2005]{Mendes05} {Mendes de Oliveira, C., Coelho, P., Gonz{\'a}lez, J.~J., \& Barbuy, B. 2005, \aj, 130, 55}
\bibitem[1995]{Menon95} {Menon, T.~K. 1995, \aj, 110, 2605}
\bibitem[1983]{Merritt83} {Merritt, D. 1983, \apj, 264, 24}
\bibitem[1984]{Merritt84} {Merritt, D. 1984, \apj, 276, 26}
\bibitem[1994]{Moles94} {Moles, M., del Olmo, A., Perea, J. et al. 1994, \aap, 285, 404}
\bibitem[2000]{Nishiura00} {Nishiura, S., Shimada, M., Ohyama, Y., Murayama, T., \& Taniguchi, Y. 2000, \aj, 120, 1691}
\bibitem[1995]{Palumbo95} {Palumbo, G.~G.~C., Saracco, P., Hickson, P. \& Mendes de Oliveira, C. 1995, \aj, 109, 1476} 
\bibitem[1994]{Paturel94} {Paturel, G., Bottinelli, L., \& Gouguenheim, L. 1994, \aap, 286, 768}
\bibitem[1997]{Paturel97} {Paturel, G., Andernach, H., Bottinelli, L., et al. 1997, \aaps, 124, 109}
\bibitem[1995]{Pildis95} {Pildis, R.~A., Bregman, J.~N., \& Evrard, A.~E. 1995, \apj, 443, 514}
\bibitem[2010a]{Plauchu10a} {Plauchu-Frayn, I. \& Coziol, R. 2010a, \aj, 139, 2643}
\bibitem[2010b]{Plauchu10b} {Plauchu-Frayn, I. \& Coziol, R. 2010b, \aj, 140, 612}
\bibitem[2004]{Proctor04} {Proctor, R.~N., Forbes, D.~A., Hau, G.~K.~T., et al. 2004, \mnras, 349, 1381}
\bibitem[1996]{Ribeiro96} {Ribeiro, A.~L.~B., de Carvalho, R.~R., Coziol, R., Capelato, H.~V., \& Zepf, S.~E. 1996, \apj, 463, 5}
\bibitem[1991]{Rubin91} {Rubin, V.~C., Hunter, D.~A., \& Ford, Jr., W.~K. 1991, \apjs, 76, 153}
\bibitem[2006]{Sanchez06} {S{\'a}nchez-Bl{\'a}zquez, P., Peletier, R.~F., Jim{\'e}nez-Vicente, J., et al. 2006, \mnras, 371, 703}
\bibitem[2007]{Schawinski07} {Schawinski, K., Kaviraj, S., Khochfar, S., et al. 2007, \apjs, 173, 512}
\bibitem[1998]{Schlegel98} {Schlegel, D.~J., Finkbeiner, D.~P., \&  Davis, M. 1998, \apj, 500, 525}
\bibitem[2006]{Skrutskie06} {Skrutskie, M.~F., Cutri, R.~M., Stiening, R., et al. 2006, \aj, 131, 1163}
\bibitem[2006]{Sulentic06} {Sulentic, J.~W., Verdes-Montenegro, L., Bergond, G., et al. 2006, \aap, 449, 937}
\bibitem[1987]{Sulentic87} {Sulentic, J.~W. 1987, \apj, 322, 605}
\bibitem[2005]{Taylor05} {Taylor, M.~B. 2005, ASPC, 347, 29}
\bibitem[2005]{Thomas05} {Thomas, D., Maraston, C., Bender, R., \& Mendes de Oliveira, C. 2005, \apj, 621, 673}
\bibitem[2010]{Thomas10} {Thomas, D., Maraston, C., Schawinski, K.,  Sarzi, M., \& Silk, J. 2010, \mnras, 404, 1775}
\bibitem[2010]{Torres10} {Torres-Flores, S., Mendes de Oliveira, C., Amram, P., et al. 2010, \aap, 521, 59}
\bibitem[2012]{TP12}{Torres-Papaqui, J.~P., Coziol, R., Ortega-Minakata, R.~A. \& Neri-Larios, D.~M. 2012, \apj, 754, 144}
\bibitem[2010]{Tzanavaris10} {Tzanavaris, P., Hornschemeier, A.~E., Gallagher, S.~C., et al. 2010, \apj, 716, 556}
\bibitem[1998]{Verdes98} {Verdes-Montenegro, L., Yun, M.~S., Perea, J., del Olmo, A, \& Ho, P.~T.~P., 1998, \apj, 497, 89}
\bibitem[2001]{Verdes01} {Verdes-Montenegro, L., Yun, M.~S., Williams, B.~A., et al. 2001, \aap, 377, 812}
\bibitem[2002]{Verdes02} {Verdes-Montenegro, L., Del Olmo, A., Iglesias-Páramo, et al. 2002, \aap, 396, 815}
\bibitem[2005]{Verdes05} {Verdes-Montenegro, L., Del Olmo, A., Yun, M.~S. \& Perea, J. 2002, \aap, 430, 443}
\bibitem[2007]{Verley07} {Verley, S., Leon, S., Verdes-Montenegro, L., et al. 2007, \aap, 472, 121}
\bibitem[1998]{Vilchez98} {V{\'{\i}}lchez, J.~M. \& Iglesias-P{\'a}ramo, J. 1998, \apjs, 117, 1}
\bibitem[2010]{Walker10} {Walker, L.~M., Johnson, K.~E., Gallagher, S.~C., et al. 2010, \aj, 140, 1254}
\bibitem[1991]{White91} {White, S.~D.~M. \& Frenk, C.~S. 1991, \apj, 379, 52}
\bibitem[1993]{Whitmore93} {Whitmore, B.~C., Gilmore, D.~M., \& Jones, C. 1993, \apj, 407, 489}
\bibitem[1987]{Williams87} {Williams, B.~A. \& Rood, H.~J. 1987, \apjs, 63, 265}
\bibitem[2005]{Yi05} {Yi, S.~K., Yoon, S.-J., Kaviraj, S., et al. 2005, \apj, 619, 111L}
\bibitem[1991]{Zepf91} {Zepf, S.~E., Whitmore, B.~C., \& Levison, H.~F. 1991, \apj, 383, 524}


\end{thebibliography}
\end{document}